

\magnification=\magstep1

\hsize = 33pc
\vsize = 46pc
\font\grand=cmbx10 at 14.4 true pt
\font\grandw=cmsy10 at 14.4 true pt

\baselineskip=12pt
\tolerance=8000
\parskip=5pt

\def\q#1{[#1]}              
\def\bibitem#1{\parindent=8mm\item{\hbox to 6 mm{\q{#1}\hfill}}}
\font\extra=cmss10 scaled \magstep0
\setbox1 = \hbox{{{\extra R}}}
\setbox2 = \hbox{{{\extra I}}}
\setbox3 = \hbox{{{\extra C}}}
\def\RRR{{{\extra R}}\hskip-\wd1\hskip2.0 true pt{{\extra I}}\hskip-\wd2
\hskip-2.0 true pt\hskip\wd1}
\def\Real{\hbox{{\extra\RRR}}}

\setbox4=\hbox{{{\extra Z}}}
\def\Z{{{\extra Z}}\hskip-\wd4\hskip 2.5 true pt{{\extra Z}}}
\def\Zed{\hbox{{\extra\Z}}}

\def\pa{\partial}
\def\P{{\cal P}}

\def\I{{\cal I}}
\def\K{{\cal K}}
\def\xp{{\hat \xi_+}}
\def\xm{{\hat \xi_-}}
\def\xpm{{\hat \xi_\pm}}
\def\J{{\hat J}}
\def\W{{\hat W}}
\def\cW{{\cal W}}
\def\G{{\cal G}}
\def\S{{\cal S}}
\def\D{{\cal D}}
\def\k{{\kappa}}
\def\R{{\cal R}}
\def\GH{\widehat{\G}}
\def\HH{\widehat{\cal H}}

\def\Gh{\hat{G}}
\def\Lh{\hat{L}}
\def\Wh{\hat{W}}
\def\hb{\hbar}

\def\slt{sl(2)}
\def\slth{\widehat{\slt}}
\def\slthoh{\widehat{\slt}_{-{1 \over 2}}}

\def\bs{\baselineskip=14.5pt}

\def\bouschou{1}
\def\bowwatts{2}
\def\dBT{3}
\def\FORT{4}
\def\fehort{5}
\def\kauwatts{6}
\def\ehh{7}
\def\hornfeck{8}
\def\ragoucy{9}
\def\bakri{10}
\def\weyl{11}
\def\howe{12}
\def\KR{13}
\def\bbss{14}
\def\gno{15}
\def\RL{16}
\def\bouwknegt{17}
\def\gosch{18}
\def\bogo{19}
\def\ralph{20}
\def\annals{21}
\def\ehhh{22}
\def\fatzam{23}
\def\BBSS{24}
\def\fatlyk{25}
\def\ravanini{26}
\def\horst{27}
\def\commute{28}
\def\FRS{29}
\def\gso{30}
\def\thielemans{31}
\def\appA{A}
\def\appB{B}

\pageno=0
\def\folio{
\ifnum\pageno<1 \footline{\hfil} \else\number\pageno \fi}

\baselineskip=12pt
\phantom{not-so-FUNNY}
\rightline{ ITP-SB--93--84\break}
\rightline{ BONN--HE--93--49\break}
\rightline{ December 1993\break}
\rightline{ hep-th/9312049\break}
\vskip 1.0truecm

\bs

\centerline
{\grand A class of {\grandw W}-algebras with
infinitely generated classical limit}

\vskip 1.0truecm
\centerline{J.\ de Boer${}^{1}$,
 L.\ Feh\'er${}^{2,}$\footnote*{An Alexander von Humboldt Fellow.
On leave from Bolyai Institute of Szeged University, H-6720 Szeged, Hungary.}
and
A.\ Honecker${}^2$
}
\bigskip
\centerline{\it ${}^{1}$Institute for Theoretical Physics}
\centerline{State University of New York at Stony Brook}
\centerline{\it Stony Brook, NY 11794-3840, USA}

\medskip
\centerline{\it ${}^{2}$Physikalisches Institut der Universit\"at Bonn}
\centerline{\it Nussallee 12, 53115 Bonn, Germany}

\vskip 1.0truecm
\centerline{\bf Abstract}
\vskip 0.2truecm
There is a relatively well understood class of deformable
${\cal W}$-algebras, resulting from Drinfeld-Sokolov (DS) type
reductions of Kac-Moody algebras, which are Poisson bracket
algebras based on finitely, freely generated rings of differential
polynomials in the classical limit.
The purpose of this paper is to point out the existence of a second
class of deformable ${\cal W}$-algebras, which in the classical limit are
Poisson bracket algebras carried by {\it infinitely, nonfreely} generated
rings of differential polynomials.
We present illustrative examples of coset constructions,
orbifold projections, as well as first class Hamiltonian reductions
of DS type ${\cal W}$-algebras leading to reduced algebras
with such infinitely generated classical limit.
We also show in examples that the reduced quantum
algebras are {\it finitely} generated
due to quantum corrections arising upon normal ordering  the
relations  obeyed by the classical generators.
We apply {\it invariant theory} to describe the relations
and to argue that classical cosets are infinitely, nonfreely generated
in general.
As a by-product, we also explain the origin of the previously constructed
and so far unexplained deformable quantum ${\cal W}(2,4,6)$ and
${\cal W}(2,3,4,5)$ algebras.

\vfill\eject

\centerline{\bf 0.\ Introduction}
\medskip

The classification of extended conformal algebras
(also called $\cW$-algebras, or local chiral algebras) is a key ingredient
to the classification of two-dimensional
rational conformal field theories, which apart
from being interesting in its own right
is also interesting since it is related to problems in
statistical physics, string theory and integrable systems.
The experience accumulated so far (see \q{\bouschou} for a review)
shows that ${\cal W}$-algebras come in two varieties.
Firstly, there exist `deformable' (or `generic') $\cW$-algebras
for which the structure constants are continuous functions
--- with isolated singularities ---
of the Virasoro centre $c$ for a fixed field content,
{\it i.e.}, for a fixed conformal weight spectrum of generating fields.
Secondly, there exist also `nondeformable' (or `exceptional')
$\cW$-algebras that appear only at particular, isolated $c$ values.
It is generally believed that, with notable exceptions,
the nondeformable $\cW$-algebras
can be understood in terms of the deformable ones,
for instance they could occur
in particular minimal models of deformable $\cW$-algebras.
Most deformable $\cW$-algebras considered so far result from
Drinfeld-Sokolov (DS) type Hamiltonian reductions of affine Kac-Moody
algebras, and thus have a classical limit which is a Poisson
bracket algebra carried by a differential polynomial ring
generated by a {\it finite} number of {\it independent} generating fields.
To put it differently,  a class of deformable
quantum $\cW$-algebras exists to which those algebras
belong which admit a {\it finitely, freely} generated classical limit.
It appears that this class of $\cW$-algebras is by now
reasonably  well understood
(see, {\it e.g.},  \q{\bowwatts--\fehort}
and references therein), though a lot of work remains to be
done before we will have it completely catalogued.

However, there are a number of reasons for believing that the above
mentioned class does not exhaust the deformable quantum $\cW$-algebras.
We have such indications in the context of each three methods usually
used for obtaining $\cW$-algebras; the direct constructions,
coset constructions and the first class Hamiltonian reduction method.
Indeed, the direct constructions provided two so far unexplained
deformable $\cW$-algebras, with conformal weights $2,4,6$
\q{\kauwatts,\ehh} and $2,3,4,5$ \q{\hornfeck},
respectively, for which the procedure used in \q{\bowwatts} for extracting
the classical limit fails. In this paper we shall explain these algebras
in terms of certain coset constructions, and shall see that their
classical analogues are in fact infinitely, nonfreely generated
(and therefore these algebras would appear to have no classical limit if one
tries to force the procedure of \q{\bowwatts} which is geared towards
a finitely generated classical limit).
Examples of classically infinitely generated
coset algebras have recently been discussed in \q{\ragoucy}.
More generally, there is a large class of coset algebras,
including for instance the diagonal cosets
$\bigl( {\widehat {\cal G}}_k \oplus \widehat {\cal G}_m\bigr)
  /{\widehat {\cal G}}_{k+m}$ at generic levels $k$ and $m$,
for which a simple group theoretic
argument given in this paper shows that
generically they have
infinitely generated  classical limits.
Finally, it is clear that the DS reductions,
which underlie all known deformable $\cW$-algebras with a finitely,
freely generated classical limit, are a very special subclass
of the reductions of Kac-Moody algebras
defined by conformally invariant  first class constraints.
In some examples of first class reductions
to which the DS mechanism does not apply
it has already been shown in \q{\FORT} that
the ring of gauge invariant differential polynomials is {\it not freely}
generated. In this paper we further develop an illustrative example of
this sort and demonstrate that the invariant ring carrying the
reduced Poisson bracket algebra is {\it infinitely, nonfreely} generated.

We wish to emphasize that although reductions of finitely generated
DS type  $\cW$-algebras and Kac-Moody algebras seem to lead generically to
{\it infinitely}, nonfreely generated algebras at the {\it classical} level,
the corresponding reduced  {\it quantum} algebras are {\it finitely}
generated in all cases studied so far\footnote{${}^{1}$}{
A claim to the contrary made in ref.~\q{\bakri}
is not correct, see also section 3.}.
We shall see that the underlying mechanism responsible for this is that the
infinitely many classical generators are not independent but obey infinitely
many relations and upon normal ordering (a subset of) the relations
quantum corrections arise which allow for eliminating the infinitely many
`would-be-generators' in favour of a finite subset.

The paper is organized as follows. We shall first analyze a very simple
example, a reduction of a $\beta$-$\gamma$ system, in detail to
illustrate the ideas. In particular, we shall see why
the quantum version of the classically infinitely generated
reduced algebra is finitely generated.
Then we shall explain how this construction is related to
a coset construction underlying the so far unexplained $\cW(2,4,6)$-algebra.
This will lead us to discussing general coset constructions as well
as bosonic projections of fermionic $\cW$-algebras and orbifolds
of $\cW$-algebras that also possess infinitely, nonfreely generated
classical analogues in general.
Finally, to show that infinitely, nonfreely generated classical algebras
arise in all reduction procedures,
we  treat an example of Hamiltonian reduction
by first class constraints that leads to such a reduced
classical system.
We give our conclusions and comment on open problems at the end of the main
text, and there are also two appendices containing technical material.

\bigskip
\centerline{\bf 1.\ A reduction of a $\beta$-$\gamma$ system}
\medskip

Below, we work out our simplest example, first at the classical
then at the quantum level.
We start by considering
two independent generating fields $\xi_\pm(z)$
defined on the circle,  subject to the Poisson brackets (PBs):
$$
\{ \xi_-(x),\xi_+(y)\}=\delta(x-y),
\qquad
\{ \xi_-(x),\xi_-(y)\}=\{\xi_+(x),\xi_+(y)\}=0.
\eqno(1.1)$$
Then we have the chiral algebra carried by the differential ring
${\cal P}$ consisting of the polynomials in $\xi_\pm^{(i)}:= \pa^i \xi_\pm$.
The fields $\xi_\pm$ are primary fields of weight ${1\over 2}$ with
respect to the conformal structure defined by
$$
T:=-{1\over 2} \bigl(\xi_+ \pa \xi_- - \xi_- \pa \xi_+\bigr).
\eqno(1.2)$$
This system is the classical version of a
linear `$\beta$-$\gamma$ system' ($\beta:=\xi_-$, $\gamma:=\xi_+$)
often used in conformal field theory.
We introduce the $\widehat{sl(2)}$ subalgebra of ${\cal P}$
generated by the currents
$$
J_H:=\xi_-\xi_+,
\qquad
J_E:=-{1\over 2} \xi_+^2,
\qquad
J_F:={1\over 2} \xi_-^2,
\eqno(1.3)$$
satisfying the PB relations
$$\eqalign{
\{ J_H(x), J_E(y)\}&=2 J_E(y)\delta,
\quad
\{ J_H(x),J_F(y)\}=-2 J_F(y)\delta,
\quad
\{ J_E(x),J_F(y)\}=J_H(y)\delta,\cr
\{ J_H(x),\xi_\pm(y)\}&=\pm \xi_\pm(y)\delta,
\quad
\{ J_E(x),\xi_-(y)\}=\xi_+(y)\delta,
\quad
\{ J_F(x),\xi_+(y)\}=\xi_-(y)\delta,\cr}
\eqno(1.4)$$
where $\delta=\delta(x-y)$.
We wish  to describe  the
reduced  chiral algebra carried by the
commutant (centralizer) ${\cal P}_{sl(2)}$  of the $sl(2)$ defined
by the zero modes of the currents (1.3) in ${\cal P}$.
In other words, we are interested in the `classical coset'
of the $\beta$-$\gamma$ algebra with respect to the `horizontal'
subalgebra
$sl(2)\subset \widehat {sl(2)} \subset {\cal P}$, {\it i.e.},
${\cal P}_{sl(2)}={\cal P}/sl(2)$.

To find the commutant notice from (1.4) that $\cal P$
is a ring of polynomials in  infinitely many variables that
form doublets,  $(\xi_+^{(i)}, \xi_-^{(i)})$ for any
$i=0,1,\ldots$, under the global $sl(2)$ transformations
generated by the zero modes of the currents.
Therefore the pairwise symplectic scalar products of the doublets,
given by
$$
W_{i,j}:=\xi_+^{(i)}\xi_-^{(j)}-\xi_-^{(i)}\xi_+^{(j)},
\eqno(1.5)$$
are obviously $sl(2)$ invariants, {\it i.e.}, belong to
the invariant subring ${\cal P}_{sl(2)}\subset {\cal P}$.
The fact that the $W_{i,j}$ are actually a {\it generating set} of
${\cal P}_{sl(2)}$ is much less obvious, but it follows from
{\it invariant theory}.
In the terminology of Weyl \q{\weyl}, this is just
the `first main theorem' of invariant theory for the
(defining representation of the) classical
group $SL(2)$.

It is perhaps worth recalling here that invariant theory
(see, {\it e.g.}, \q{\weyl,\howe}) deals with the following class
of problems (among others). Take a group $G$ and a set
$V_\alpha,  \ldots , V_\omega$ of finite dimensional
representations of $G$.
Consider the space ${\cal M}$ of  polynomials $p$
depending on (a fixed or arbitrary number of) variables
that belong to these representations,
$p=p(v_\alpha^0,\ldots, v_\alpha^{m_\alpha}, \ldots,v_\omega^0,\ldots,
v_\omega^{m_\omega})$, where $v_\alpha^{i_\alpha}\in V_\alpha$
for $i_\alpha=0,\ldots, m_\alpha$ etc.
The linear space ${\cal M}$ carries a natural representation of $G$
induced by transforming the arguments of $p$.
The problem then is to describe the $G$-singlets, {\it i.e.},
the invariant polynomials.
An important point is that since ${\cal M}$ is a polynomial
ring so is the subring of invariants.
The two major questions are:
i) Describe a generating set of the invariant ring.
ii) Find the relations obeyed by the (in general not algebraically
    independent) generators.
The answer to i) is called the
`first main theorem' and the answer to ii) the `second
main theorem' of invariant theory for a given problem.
In particular, in \q{\weyl} these problems are solved for $G$ a
classical group and ${\cal M}$ the space of polynomials depending on
an arbitrary number of variables in the vector (defining)
representation of $G$.

Returning to  our problem, it is easy to check that in addition to
antisymmetry,
$$
W_{i,j}+W_{j,i}=0,
\eqno(1.6)$$
the generators $W_{i,j}$ satisfy the following relation:
$$
W_{i,j}W_{k,l} - W_{i,k}W_{j,l}+W_{i,l}W_{j,k}=0,
\eqno(1.7)$$
for any $i,j,k,l$.
The relation (1.7) is known as the `syzygy' in invariant theory, and
the `second main theorem' of invariant theory  states
that all polynomial relations between the $sl(2)$ invariant
polynomials  are consequences of (1.6) and (1.7)
(see Chapter VI.1 in \q{\weyl} and also  Corollary 3.2 in \q{\KR}).
Of course, this statement holds if we consider the variables $\xi_\pm^{(i)}$
independent, {\it i.e.}, if we forget about the action of the derivation
$\pa$ of the differential ring $\cal P$, given by
$\pa \xi_{\pm}^{(i)}=\xi_\pm^{(i+1)}$.
Taking the derivation into account we also have the relation
$$
\pa W_{i,j}=W_{i+1,j}+W_{i,j+1}.
\eqno(1.8)$$
The generating set $\{ W_{i,j}\}$ is overcomplete (redundant) on account
of (1.8) and to describe the reduced PB algebra in the most economical way
we should  select a subset of the generators
forming a minimal generating set of ${\cal P}_{sl(2)}$.
For this consider the linear span of the redundant set of
generators at scale dimension $d$:
$$
{\cal V}_d:=\hbox{ linear span}\{ W_{i,j}\,\vert\, i+j+1=d\,\}.
\eqno(1.9)$$
The derivation $\pa$ maps ${\cal V}_d$ into ${\cal V}_{d+1}$
according to (1.8) and it is easy to see that by using this we
can express the generators $\{\, W_{i,j}\,\vert\, \forall\, i,j\,\}$
in terms   of a subset of the generators
spanning one dimensional subspaces of ${\cal V}_{2r}$ for
$r=1,2,\ldots$ any natural number.
In this way we obtain a nonredundant (minimal) generating set of
${\cal P}_{sl(2)}$, for instance $\{ W_{1,2s}\}$ with $s$ running
over the nonnegative integers. In terms of the redundant generating
set the reduced chiral algebra is given by
$$\eqalign{
\{ W_{i,j}(x),W_{k,l}(y)\}=&
(-1)^{i+1}\sum_{a=0}^{i+l}
   (-1)^a {i+l\choose a} W_{k,i+j+l-a}(y)\delta^{(a)}(x-y)\cr
+&(-1)^{j}\sum_{b=0}^{j+l}
   (-1)^b {j+l\choose b} W_{k,i+j+l-b}(y)\delta^{(b)}(x-y)\cr
+&(-1)^{j+1}\sum_{c=0}^{j+k}
   (-1)^c {j+k\choose c} W_{l,i+j+k-c}(y)\delta^{(c)}(x-y)\cr
+&(-1)^{i}\sum_{d=0}^{i+k}
   (-1)^d {i+k\choose d} W_{l,i+j+k-d}(y)\delta^{(d)}(x-y).\cr}
\eqno(1.10)$$
In the present example the r.h.s.\ is linear
in the generators and we observe that
this feature is also
valid for the PB's of the generators $\{ W_{1,2s}\}$, as the elements
of the redundant generating set are given by linear expressions
in terms of the nonredundant generating set $\{ W_{1,2s}\}$.
The elements in the nonredundant generating set $\{ W_{1,2s}\}$ are of course
not independent, since the number of degrees of freedom should not increase
in a reduction and we started with just two generating fields $\xi_\pm$.
Indeed, they satisfy the infinitely many differential--algebraic relations
that can be obtained from (1.6), (1.7) and (1.8) by expressing all
$W_{i,j}$ in terms of $\{ W_{1,2s}\}$.

The main result of the above analysis is that
{\it infinitely many generators obeying infinitely many
differential--algebraic relations} are needed to describe the reduced
classical chiral algebra carried by ${\cal P}_{sl(2)}$.
We next analyze this reduction at the quantum level.
For this we take two bosonic chiral quantum fields
$\xp(z)$, $\xm(z)$, where the argument now varies on the
punctured complex plane, for which the only nonregular OPE is
$$
\xm(x)\xp(y):={\hbar \over {x-y}}+ {\rm reg}.
\eqno(1.11)$$
Here we introduced the Planck constant $\hbar$ explicitly in order
to make clear the correspondence to the PB (1.1), but one could
of course set $\hbar$ to unity by a rescaling of the fields.
The fields $\xpm$ are primary fields of weight ${1\over 2}$ with
respect to the conformal structure defined by
$$
\hat T:=-{1\over 2} \bigl(\xp \pa \xm - \xm \pa \xp\bigr).
\eqno(1.12)$$
We adopted the notation of Bais {\it et al} \q{\bbss}, thus $\bigl( AB\bigr)$
denotes the usual {\it normal ordered} product of the fields $A$ and $B$.
We also define the quantum $\widehat{sl(2)}$ currents by normal
ordering the expressions in (1.3),
$$
\J_H:=\bigl(\xm\xp\bigr),
\qquad
\J_E:=-{1\over 2} \bigl(\xp\xp\bigr),
\qquad
\J_F:={1\over 2} \bigl(\xm\xm\bigr),
\eqno(1.13)$$
and find the (only nonregular) OPE's of the quantum currents to be
$$\eqalign{
\J_H(x) \J_E(y)&={2\hbar \J_E(y)\over x-y} +{\rm reg.}\cr
\J_H(x) \J_F(y)&={-2\hbar \J_F(y)\over x-y} +{\rm reg.}\cr
\J_H(x) \J_H(y)&={-{\hbar}^2 \over (x-y)^2} +{\rm reg.}\cr
\J_E(x) \J_F(y)&=-{1\over 2}{{\hbar}^2\over (x-y)^2} +
{\hbar \J_H(y)\over x-y} +{\rm reg.}\, , \cr}
\eqno(1.14)$$
which (by putting $\hbar =1$) is $\widehat{sl(2)}$ at level $-{1\over 2}$.
The fields $\xpm$ generate the chiral algebra $\hat{\cal P}$,
{\it i.e.}, the linear space $\hat {\cal P}$
is spanned by  the (repeated) derivatives and normal ordered
products of $\xpm$. The object of our interest is the coset
$$
{\hat {\cal P}}_{sl(2)}:={{\hat {\cal P}}\over sl(2)},
\eqno(1.15{\rm a})$$
{\it i.e.},
the set of fields in $\hat {\cal P}$ that commute
with  the horizontal subalgebra $sl(2)\subset {\widehat {sl(2)}}
\subset \hat {\cal P}$ spanned by the charges
$$
Q_A={1\over 2\pi i} \oint dx \J_A(x),
\qquad A=E,F,H.
\eqno(1.15{\rm b})$$
It is straightforward to verify that the quantum versions of the invariants
in (1.5), given by
$$
\W_{i,j}:=\bigl(\xp^{(i)}\xm^{(j)}\bigr)
         -\bigl(\xm^{(i)}\xp^{(j)}\bigr),
\qquad  \xpm^{(i)}:= \pa^i \xpm,
\eqno(1.16)$$
belong to ${\hat {\cal P}}_{sl(2)}$.
 In the classical limit $\hat {\cal P}_{sl(2)}$
becomes the differential ring ${\cal P}_{sl(2)}$ and
we know that the vector space ${\cal P}_{sl(2)}$ is spanned
by the (repeated) derivatives and products of the $W_{i,j}$.
{}From this we can conclude that
$\hat {\cal P}_{sl(2)}$
is spanned by the derivatives and normal ordered products of
the $\W_{i,j}$.
The OPE's of the generating set $\{\W_{i,j}\}\subset{\hat{\cal P}}_{sl(2)}$
read
$$\eqalign{
\W_{i,j}(x) \W_{k,l}(y)=&
(-1)^{i+1}\hbar \sum_{a=0}^{i+l}
  {i+l\choose a} \W_{k,i+j+l-a}(y)\pa_y^a{1\over x-y}\cr
+&(-1)^{j}\hbar \sum_{b=0}^{j+l}
  {j+l\choose b} \W_{k,i+j+l-b}(y)\pa_y^b{1\over x-y}\cr
+&(-1)^{j+1}\hbar \sum_{c=0}^{j+k}
  {j+k\choose c} \W_{l,i+j+k-c}(y)\pa_y^c{1\over x-y}\cr
+&(-1)^{i}\hbar \sum_{d=0}^{i+k}
  {i+k\choose d} \W_{l,i+j+k-d}(y)\pa^d_y{1\over x-y}\cr
+&2{\hbar}^2{(-1)^{i+j+1}[(i+l)!(j+k)!
   -(i+k)!(j+l)!]\over (x-y)^{i+j+k+l+2}}+{\rm reg}.\cr}
\eqno(1.17)$$
Observe that, upon the correspondence
$\pa_y^n {1\over x-y} \longleftrightarrow \pa_y^n \delta(x-y)$,
the OPE (1.17) corresponds to the PB (1.10), with a quantum correction
given by the last ${\cal O}({\hbar}^2)$ term.
 In particular, noting that $\hat T = {1\over 2} \W_{1,0}$,
we can confirm that the central charge of our system is $c=-1$,
which is a special case of a well-known result about
$\beta$-$\gamma$ systems.

We now come to the main point, namely,
the implication of the normal ordered version of the syzygy (1.7).
A straightforward calculation, based on the normal ordering
rearrangement identities given, for instance,
in  \q{\bbss}, leads to the following result:
$$\eqalign{
&\left( \W_{i,j} \W_{k,l}\right) - \left( \W_{i,k}\W_{j,l}\right)
+\left(\W_{i,l}\W_{j,k}\right) =\cr
&{\hbar} \left[ C^j_{i,k,l} \W_{j,i+k+l+1} + C^k_{i,l,j} \W_{k,i+j+l+1}
+C^l_{i,j,k} \W_{l,i+j+k+1}\right] ,\cr}
\eqno(1.18{\rm a})$$
with linear combination coefficients given by
$$
C^l_{i,j,k}= \Bigl[
{\left((-1)^k+(-1)^{j+1}\right) (j+k)!\over (j+k+1)!}
+{(-1)^i (i+k)!\over (i+k+1)!} + {(-1)^{i+1} (i+j)!\over (i+j+1)!}
\Bigr].
\eqno(1.18{\rm b})$$
Similarly to the classical case, by using only
$\pa \W_{a,b}=\W_{a+1,b}+\W_{a,b+1}$
we can express the generators $\{\, \W_{i,j}\,\vert\, \forall\, i,j\,\}$
in terms of a subset of the generators
spanning one dimensional subspaces of ${\hat {\cal V}}_{2r}$,
$$
{\hat {\cal V}}_d:=\hbox{ linear span}\{ \W_{i,j}\,\vert\, i+j+1=d\,\},
\eqno(1.19)$$
for $r=1,2,\ldots$ any natural number.
However, since the r.h.s.\ of (1.18a) is nonzero for $\hbar\neq 0$,
it is clear that using also (1.18a) we can now eliminate
the remaining generators lying in  ${\hat {\cal V}}_{2r}$ as well if
$2r$ can be written in the form
$2r=i+j+k+l+2$
with pairwise distinct nonnegative integers $i,j,k,l$.
But this is always possible if
$2r\geq 0+1+2+3+2=8$,
which implies that the commutant $\hat \P_{sl(2)}$
is generated by (repeated) derivatives and normal
ordered products of the following {\it finite} set of fields:
$$
\hat T={1\over 2} \W_{1,0},
\quad
\W_{1,2},
\quad
\W_{1,4},
\eqno(1.20)$$
having  scale dimensions $2$, $4$, $6$.
This is a drastic difference  from  the classical
case where we need infinitely many generating fields since
the r.h.s.\ of  the syzygy (1.7)
is zero in the classical case, when $\hbar =0$.
The (automatically closed, nonlinear) OPE algebra of the fields (1.20)
can be found from (1.17).
It is also easy to find a generating set of the commutant
$\hat \P_{sl(2)}$ consisting of $\hat T$ and {\it primary} fields
with weights $4$ and $6$.
In conclusion, the above construction yields a quantum ${\cal W}(2,4,6)$
algebra at $c=-1$
possessing an infinitely, nonfreely generated classical limit.
Next  we shall identify this $\cW$-algebra
as the coset of ${\widehat {sl(2)}}$ at level $-{1\over 2}$
with respect to the horizontal subalgebra.

In order to establish the  isomorphism
given by the second equality in
$$
\hat \P_{sl(2)}:={{\hat {\cal P}}\over {sl(2)}}=
{{\widehat{sl(2)}}_{-{1\over 2}}\over sl(2)},
\eqno(1.21)$$
where
${{\widehat{sl(2)}}_{-{1\over 2}} / sl(2)}$
denotes the commutant of the horizontal $sl(2)$ in
${\widehat{sl(2)}}_{-{1\over 2}}\subset \hat {\cal P}$,
it is enough to verify that the generators
of $\hat \P_{sl(2)}$ given by (1.20) can be
expressed in terms of the $\widehat{sl(2)}_{-{1\over 2}}$ currents
given by (1.13).
The quantum Sugawara Virasoro of the current
algebra is given by
$$
3\hbar {\hat T}_{\rm sug}=\bigl(\hat J\cdot \hat J\bigr)=
{1\over 2} \bigl(\J_H \J_H\bigr) + \bigl( \J_E \J_F\bigr)
   +\bigl(\J_F \J_E\bigr).
\eqno(1.22)$$
Using the rearrangement identities of \q{\bbss},
one can prove the equality
$$
\hat T_{\rm sug} =  \hat T,
\eqno(1.23)$$
where $\hat T$ is given by (1.12). Note that the l.h.s.\ of
(1.23) is quartic in the basic fields $\xpm$,
while the r.h.s.\ is quadratic (reminding  us of
the ``symmetric space theorem'' \q{\gno}).
Clearly, such an equality is only possible at
the quantum level, and in fact the classical analogue of the
Sugawara expression (1.22) vanishes identically.

Concerning the weight $4$ and the weight $6$ generators,
we can establish the following identities:
$$
\bigl(\pa \hat J \cdot \pa \hat J\bigr)
 = {1\over 2 }\bigl(  \W_{1,0}  \W_{1,0}\bigr)
 -\hbar \bigl[ {1\over 6} \pa^2 \W_{1,0} +
 {5\over 3} \W_{1,2}\bigr],
\eqno(1.24)$$
and
$$
\bigl(\pa^2 \hat J \cdot \pa^2 \hat J\bigr)
={1\over 2} \bigl(\pa \W_{1,0}\, \pa \W_{1,0}\bigr)
 -2\bigl( \W_{1,0} \W_{1,2}\bigr)
 -\hbar \bigl[ {1\over 60} \pa^4 \W_{1,0} + {38\over 15} \pa^2 \W_{1,2}
 -{17\over 5} W_{1,4}\bigr].
\eqno(1.25)$$
It is easy to see that using the above identities
one can express the generating set
(1.20) of $\hat \P_{sl(2)}$ in terms of composites of the current
$\hat J$. (Observe that this is possible because quantum
corrections are present on the r.h.s.\ of (1.24) and (1.25)).
This proves the relation $\hat \P_{sl(2)} \subset
{{\widehat{sl(2)}}_{-{1\over 2}} / sl(2)}$. Since we also have
${\slth_{-{1\over 2}} / sl(2)} \subset {\hat \P}_{sl(2)}$
on account of $\slth_{-{1\over 2}}\subset \hat {\cal P}$, by (1.13),
the isomorphism  (1.21)  has been now established.

\bigskip
\centerline{\bf 2.\ The deformable $\cW(2,4,6)$ and
                remarks on diagonal cosets}
\medskip

In this section we  explain the origin of the
`fourth' deformable $\cW(2,4,6)$-algebra, which was
unexpectedly found in \q{\kauwatts} in addition to the
3 expected deformable algebras with the same weights
(the DS type $\cW_{\cal S}^{\cal G}$-algebras
--- or `Casimir algebras' ---
corresponding to the principal $sl(2)$ embedding in $B_3$, $C_3$ and
the bosonic projection of the $N=1$ super Virasoro algebra).
The explanation will be given in terms of a  coset construction.
We also present related general considerations on coset constructions
of $\cW$-algebras, arguing that {\it in the  classical case}
the diagonal cosets  generically yield
infinitely, nonfreely  generated algebras.

One can calculate the vacuum character, $\chi_0(q)$, of the $\cW(2,4,6)$
constructed in the previous section at $c=-1$ from either of its coset
realizations in (1.21).
One obtains (see also \q{\RL}) the following formula:
$$
\chi_0(q)={{\varphi_0(q)-q\varphi_2(q)}\over \prod_{n>0} (1-q^n)^2}
\qquad\hbox{with}\qquad
\varphi_n(q):=\sum_{m\geq 0} (-1)^m q^{{m(m+1)\over 2}+mn}\,.
\eqno(2.1)$$
If one compares this with
the standard `vacuum Verma module character'  $\phi_{2,4,6}(q)$
associated to fields with weights $2$, $4$, $6$,  given by
$$
\phi_{2,4,6}(q):={1\over \prod_{n\geq 0} (1-q^{n+2})
(1-q^{n+4})(1-q^{n+6})}\,,
\eqno(2.2)$$
one finds up to order 20
$$
\chi_0(q)-\phi_{2,4,6}(q)=
- q^{11} (1 + 2 q + 3 q^2 + 6 q^3 + 10 q^4 + 17 q^5 + 27
q^6 + 44 q^7 + 67 q^8 + 105 q^9 + {\cal O}(q^{10})).
\eqno(2.3)$$
The fact that all coefficients on the r.h.s.\ of (2.3)  are nonpositive
and $\phi_{2,4,6}(q)$ is the smallest standard character for which this
is the case not only confirms (by the argument of \q{\bouwknegt,\bouschou})
that the coset (1.21) should be a ${\cal W}(2,4,6)$-algebra, what
we have already proved,
but also shows the existence of a first `null field' at scale dimension 11.
The existence of this null field is due to the normal ordered syzygy (1.18).
Indeed, at scale dimension 11 eq.~(1.18a) yields 3 relations,
correspondingly to the 3 different possible choices of pairwise distinct
$i,j,k,l$ such that $i+j+k+l+2=11$. Two linear combinations of these
relations arise as derivatives of the relations at scale dimensions $8$ and
$10$ which were used to express the scale dimension  $8$ and $10$
`would-be-generators' ($\W_{1,6}$ and $\W_{1,8}$) in terms of the set (1.20).
The remaining third relation gives rise to the null field in question.
Along these lines, it is also easy to derive  the explicit formula of this
vanishing nontrivial normal ordered differential polynomial in the generators
(1.20), but the formula is not particularly enlightening.
What is important is to emphasize the twofold r\^ole of the normal ordered
relations:
A subset of them  is responsible for the algebra being finitely
generated at the quantum level and the rest give rise to null fields.

Let us recall now that it was shown in \q{\ehh} that the so far unexplained
algebra with spins 2, 4 and 6 has a generic null field precisely
at this scale dimension.
It is also known that  the bosonic projection of the
$N=1$ super Virasoro algebra has a generic null field already at scale
dimension 10 \q{\bouwknegt}, whereas the Casimir algebras of $B_3$,
$C_3$ have no null fields at $c=-1$.
In conclusion, we can identify the coset algebra (1.21)
as the `fourth' deformable $\cW(2,4,6)$-algebra at $c=-1$.
 In order
to understand  how to deform this coset algebra to generic $c$ and thus
completely explain the algebra, let us proceed with some general
remarks on cosets.

Consider a coset of the form $\bigl( {\widehat {\cal G}}_k \oplus
\widehat {\cal G}_m\bigr)/{\widehat {\cal G}}_{k+m}$, where $\G$
is a simple Lie algebra, ${\widehat {\cal G}}_k$ is
the corresponding affine Kac-Moody algebra at level $k$,
and  $\widehat{{\cal G}}_{k+m}\subset \widehat{{\cal G}}_k \oplus
\widehat{{\cal G}}_m$ is the diagonal embedding.
In such a situation one can show that
$$\lim_{m \to \infty}
{ {\widehat {\cal G}}_k \oplus \widehat {\cal G}_m \over
  {\widehat {\cal G}}_{k+m}}
= {{\widehat {\cal G}}_k \over {\cal G}}\,,
\eqno(2.4)$$
where the equality holds at the level of algebras.
For generic $m$,  the coset algebra on the l.h.s.\
is a deformation of the coset algebra on the r.h.s.,
in particular, it has the same spin content of generators.
This statement holds for classical as well as quantum $\cW$-algebras.
The argument presented for the quantum case in \q{\gosch,\bogo}
is roughly the following: The coset generators of the l.h.s.\ are
${\cal G}$-singlets in ${\widehat {\cal G}}_k \oplus \widehat {\cal G}_m$
with respect to the horizontal subalgebra
${\cal G}\subset {\widehat {\cal G}}_{k+m}$.
Denote the currents generating ${\widehat {\cal G}}_k$ by $J_a$
($a=1,\ldots ,{\rm dim\,}\G$),
those generating $\widehat{\cal G}_m$ by $j_a$.
Taking commutators (resp.\ Poisson brackets)
with all $(J_a + j_a)$ one concludes from the central terms
that the coefficient of a monomial contained in a generator of
the coset algebra tends to zero
for $m \to \infty$ if it contains any $j_a$. This proves the inclusion of
the l.h.s.\ into the r.h.s\ of (2.4). The equality in (2.4) as well as
deformability follow from the fact that to any ${\cal G}$-singlet
consisting of the $J_a$ only one can add correction terms containing
also the $j_a$ such that it commutes with ${\widehat {\cal G}}_{k+m}$
\q{\gosch,\bogo}.

In the classical case the deformation of the `singlet algebra' on
the r.h.s.\ of (2.4) can be made explicit as follows.
The $\G$-valued current $J$ generating the algebra
$\widehat{\cal G}_k$ satisfies the Poisson brackets
$$
\{ J_a(x), J_b(y)\} =
    \sum_c f_{ab}^c \ J_c(y)\delta(x-y) - k \, g_{ab} \delta'(x-y),
\eqno(2.5)$$
where $f_{ab}^c$ ($g_{ab}$) denote  the structure constants (metric)
of $\G$, and similarly for the currents $j$ and $(J+j)$
that generate ${\widehat{\G}}_m$ and ${\widehat{\G}}_{k+m}$, respectively.
Let $P=P(J, \pa J,\ldots, \pa^n J)$ be an arbitrary element of the
singlet algebra, {\it i.e.}, a differential polynomial in the
components of $J$ which is invariant under the transformation
$\delta_\epsilon  J = [\epsilon, J]$ for any {\it constant}
$\epsilon \in \G$. The corresponding element of the diagonal
coset on the l.h.s.\ of (2.4) is obtained  by replacing $J$ by $I$, where
$$
I:= J -{k\over m} j\,,
\eqno(2.6{\rm a})$$
and also replacing $\pa^i J$ by $\D^i I$, where $\D$ is the covariant
derivative defined by
$$
\D I:= \pa I+{1 \over k+m} [J+j, I],
\eqno(2.6{\rm b})$$
for $k$, $m$, $(k+m)$ nonzero.
Indeed, for any $\G$-valued function $\epsilon(x)$,  under the transformation
$$
\delta_\epsilon J = [\epsilon, J] - k\, \epsilon'\,,
\qquad
\delta_\epsilon j = [\epsilon, j] - m\, \epsilon'\,
\eqno(2.7{\rm a})$$
one has
$$
\delta_\epsilon \left( \D^i I\right)=[\epsilon, \D^i I].
\eqno(2.7{\rm b})$$
This immediately implies that $P(I, \D I,\ldots, \D^n I)$ belongs to the
coset on the l.h.s.\ of (2.4), and it tends to $P(J,\pa J,\dots, \pa^n J)$
as $m\to \infty$, in accordance with  (2.4).

We shall see later in this paper that
covariant derivatives  are  useful also in  many other
considerations concerning current algebras,
simply because the currents generate gauge  transformations.
See also ref.~\q{\ragoucy}, which inspired some of our considerations.

We now wish to argue that the cosets ${\widehat {\cal G}}_k /{\cal G}$
are {\it classically} always {\it infinitely} generated with {\it infinitely}
many relations.
To see this we note first that invariant
theory ({\q{\weyl,\howe} and references therein) applied to the
${\cal G}$-invariant differential polynomials in the
${\cal G}$-valued current $J$
containing the derivatives $\pa^nJ$ up to a fixed finite order
leads to finitely many generators and a
finitely generated set of relations.
However,  the coset on the r.h.s.\ of (2.4) consists of
the ${\cal G}$-invariant differential
polynomials with an arbitrary number of derivatives.
Thus, in the end, one is looking for invariant polynomials
in  infinitely many variables, $\pa^nJ$ for any $n$,
and therefore the invariant ring is also generated by
infinitely many generators and infinitely many relations.
In the spirit of Weyl \q{\weyl}, these would be obtained by inserting the
infinitely many variables $\pa^n J$ into some finite list of
`typical basic invariants' and a corresponding finite list of
`typical basic relations', which would be given by a first and
a second main theorem of invariant theory  for the adjoint
representation  of ${\cal G}$ on ${\cal G}$.
(Unfortunately, we could not find these
theorems in the mathematical literature for the
adjoint representation in general, but
it might be possible to infer the case of the adjoint
representation from more general results and constructions
of invariant theory).
More precisely, since in  our context the infinitely
many variables $\pa^n J$ are linearly related
by the action of the derivation $\pa$,
using $\pa$  one could always
write down additional linear relations between those generators differing
only in the places where the derivatives have been inserted (like in (1.8)).
However, taking all relations into account one is still left with
infinitely many generators (and relations).
This argument shows that in the classical case the coset on the r.h.s.\
of (2.4) is infinitely generated with infinitely many relations.
Clearly, the same is true for the diagonal cosets
$\bigl( {\widehat{\cal G}}_k \oplus \widehat{\cal G}_m\bigr)
  /{\widehat{\cal G}}_{k+m}$ by deformability.

Let us illustrate the above in the simplest nontrivial case, the classical
singlet algebra
$$
{{\widehat{sl(2)}}_k\over sl(2)}.
\eqno(2.8)$$
Since in the complex case the adjoint representation of the Lie algebra
$sl(2)$ is equivalent to the vector representation of $o(3)$,
we can directly apply the results in \q{\weyl} for describing the generators
and relations of the ring of invariant polynomials (2.8).
More precisely, we can do this  provided we forget about the action
of the derivation and, for the moment, consider the variables
$\pa^n J$ as independent, where $J$ is the $sl(2)$ valued current.
According to \q{\weyl},
the invariant polynomials are generated by the
quadratic invariants
$$
S(m,n):=\langle \pa^m J, \pa^n J\rangle
\qquad
(\forall\, m,n),
\eqno(2.9{\rm a})$$
where $\langle\cdot,\cdot\rangle$ is given by the trace
in the defining representation of $sl(2)$,
and the cubic invariants
$$
S(p,q,r):=\langle [\pa^p J , \pa^q J], \pa^r J\rangle
\qquad
(\forall\,p\neq q\neq r).
\eqno(2.9{\rm b})$$
(This is familiar from many physical applications of the rotation group
$SO(3)$, where (2.9a) becomes the usual scalar product of two
vectors, and (2.9b) the volume of the parallelepipedon spanned by
three vectors).
We now quote the  basic relations from \q{\weyl} (Chapter II.17).
Let  $\{ m_0, m_1, m_2, m_3\}$ be any set of distinct, nonnegative integers,
and let $\{ n_0, n_1, n_2, n_3\}$ be another set of this type
(there could be overlap between the two sets).
In addition to the relations expressing the obvious symmetry properties
of the invariants (2.9), the  basic relations are of the following three
types. First,
$$
\det \left[\matrix{
S(m_0,n_0)&S(m_0,n_1)&S(m_0,n_2)&S(m_0,n_3)\cr
S(m_1,n_0)&S(m_1,n_1)&S(m_1,n_2)&S(m_1,n_3)\cr
S(m_2,n_0)&S(m_2,n_1)&S(m_2,n_2)&S(m_2,n_3)\cr
S(m_3,n_0)&S(m_3,n_1)&S(m_3,n_2)&S(m_3,n_3)\cr}\right]=0.
\eqno(2.10{\rm a})$$
Second,
$$
S(m_1,m_2,m_3) S(n_1,n_2,n_3) +2
\det \left[\matrix{
S(m_1,n_1)&S(m_1,n_2)&S(m_1,n_3)\cr
S(m_2,n_1)&S(m_2,n_2)&S(m_2,n_3)\cr
S(m_3,n_1)&S(m_3,n_2)&S(m_3,n_3)\cr}\right]=0.
\eqno(2.10{\rm b})$$
Third,
$$
\sum_m\, \pm\, S(m_1,m_2,m_3) S(m_0,n) =0\,,
\eqno(2.10{\rm c})$$
where $n$ is arbitrary and the sum is over the (signed) permutations of
$\{m_0,m_1,m_2,m_3\}$.
These would generate all relations between the invariants if the
variables $\pa^n J$ were independent.
Taking the action of the derivation into account, we
also have the linear relations
$$
\pa S(m,n)=S(m+1,n) + S(m,n+1)
\eqno(2.11{\rm a})$$
and
$$
\pa S(p,q,r)=S(p+1,q,r)+S(p,q+1,r)+S(p,q,r+1),
\eqno(2.11{\rm b})$$
where the cubic invariant is of course zero if any of its two arguments
coincide.

We can use the linear relations in (2.11),
which are analogous to (1.8),
 to introduce a {\it nonredundant}
(minimal) generating set of the differential ring of singlets (2.8).
Let $N_s$ be the number of generators at scale dimension $s$
in the minimal generating set.
{}Fom (2.11), we find the
(classical) generating function, $f_{\rm cl}(u):=\sum_s N_s u^s$,
to be given by
$$
f_{\rm cl}(u) =
{(u^6 - u^5 + u^2)\over (1-u^2) (1-u^3)}
=u^2 + u^4 + 2 u^6 +2 u^8 + u^9 + 2 u^{10} + u^{11} + 2 u^{12} + u^{13}
+3u^{14} + {\cal O}(u^{15}).
\eqno(2.12)$$
This infinite spectrum
of classical generating fields\footnote{${}^{2}$}{Eq.~(2.9) defines
invariants  for any $\G$ and
one can confirm already from the quadratic invariants
that the classical coset (2.4) is always infinitely generated.}
 is to be contrasted
with the corresponding quantum case.
We recall \q{\ralph,\bouschou} that, according to character arguments,
the quantum version of the coset (2.8) is expected to yield a
{\it finitely} generated algebra. Moreover, for {\it generic} $k$
the scale dimensions of the quantum generating
fields should be those determined by the following (quantum)
generating function, $f_q(u)$,
$$
f_q(u)= u^2 + u^4 + 2 u^6 +2 u^8 + u^9 + 2 u^{10} +  u^{12}.
\eqno(2.13)$$
The first difference between the classical and quantum generating functions
occurs at scale dimension $11$, where we have a classical generator but no
quantum one in the respective nonredundant generating sets.
The explanation lies in the fact that the first classical relation also
occurs at precisely this scale dimension, namely, it is given by (2.10c)
with $\{ m_0, m_1, m_2, m_3\}=\{ 0,1,2,3\}$ and $n=0$.
Given our experience with the $\beta$-$\gamma$ example, we now expect the
following:
A redundant, infinite  generating set of the quantum coset is obtained by
normal ordering the classical generating fields, and for the quantum coset it
must be possible to eliminate the `would-be-generator' at scale dimension
$11$ due to a quantum correction in the normal ordered version of (2.10c).
In a similar fashion, one can also understand the cancellation of
one `would-be-generator' at scale dimension 12.
The statement that the quantum generating function (2.13) is correct,
which is supported but not rigorously
proved by the  character `argument',   is clearly equivalent
to this cancellation taking place at all higher scale
dimensions\footnote{${}^{3}$}{
Similarly to the $\beta$-$\gamma$ example,
only a subset of the relations should be  needed for the cancellation,
the rest should give rise to generic --- in $k$ --- null fields.
In fact, the first such generic null fields appears at scale dimension $13$.
}.

The coset
$${\slth_k \oplus \slth_m \over \slth_{k+m}}
\eqno(2.14)$$
can now be treated easily: We combine the general remarks on cosets
of type (2.4) with the statements on the coset $\slth_k / \slt$.
Classically, a generating set for the coset algebra (2.14) can be
obtained as follows. In the generators (2.9) one replaces the current
$J$ by the current $I$ as given by (2.6a) and substitutes the derivative
$\pa$ by the covariant derivative $\D$ (2.6b). The relations in the
coset (2.14) are obtained by the same substitutions applied to
the relations (2.10), (2.11) in the coset $\slth_k / \slt$.

The quantum version of the coset (2.14) is more complicated because,
due to the covariant derivative $\D$, the generators of the
quantum algebra cannot be obtained by naively normal ordering
the classical generators. At least, deformability
ensures that for generic $m$
we have the same number of generators and null fields
in the quantum coset $\slth_k \oplus \slth_m / \slth_{k+m}$
as in
the coset $\slth_k / \slt$. In particular, eq.\ (2.13) means
that the quantum coset (2.14) leads to a
$\cW(2,4,6,6,8,8,9,10,10,12)$
(with known truncations for integer positive $m<6$
\q{\ralph,\bouschou}).
Still, the results from invariant theory also simplify the
explicit construction of the generators of this quantum coset.
Noting that
$\slth$ is generated by the two subalgebras $\slt$ and $\widehat{U(1)}$
one may restrict the ansatz for invariant fields to
normal ordered differential polynomials
in the basic invariants (2.9) where each $\slt$ valued current
$J$ can either be replaced by $J \in \slth_k$ or by $j \in \slth_m$.
Then, one determines the coefficients in this ansatz by
requiring the OPE with the $U(1)$-current in the diagonally embedded
$\slth$ to be regular. This enables one to explicitly construct at
least the primary generator of dimension 4 in the coset (2.14)
in addition to the Virasoro field.

Before proceeding with the explanation of the deformable quantum
$\cW(2,4,6)$, we would like to note a well-known result for the
central charge $c$ of the quantum coset (2.14):
$$c_k(m) = {3 k \over k+2} \left(1- {2 (k+2) \over (k+m+2) (m+2)} \right).
\eqno(2.15)$$

Now we return to the starting point of this section and give the
explanation of the `fourth' deformable $\cW(2,4,6)$. Recall that
we have already identified this algebra at $c=-1$ as the coset
$\slthoh / \slt$.
Eq.~(2.4) enables us to deform the quantum coset $\slthoh / \slt$
to generic $c$ (respective $m$) in the following manner
\footnote{${}^{4}$}{
This observation is due to R.\ Blumenhagen.
}:
$${\slthoh \oplus \slth_m \over \slth_{m-{1 \over 2}}} \cong  \cW(2,4,6).
\eqno(2.16)$$
In particular, also the null field at scale dimension 11 is deformed
to generic $c$. This shows that the coset (2.16) realizes the previously not
understood solution for $\cW(2,4,6)$. The relation between the central
charge $c$ and the level $m$ is given by (2.15): $c = c_{-{1 \over 2}}(m)
= -{(2 m +7) m \over (2 m +3) (m+2)}$.
(We have not assumed $m$ to be an integer and therefore we have
indeed constructed the algebra for {\it generic} $c$).
When looking for minimal models of this $\cW(2,4,6)$ particular
values of the level $m$ will be distinguished.
Note that, by an explicit search,
a few minimal models of this algebra  have been found in \q{\ehh}
for values of
$c$ where the level $m$ is an integer. It would certainly be
interesting to find out if this is true in general and to better
understand the minimal models of the coset algebra (2.16).

We conclude the discussion of the example (2.16) by remarks on
how the truncation happens at $k=-{1 \over 2}$ in the generically larger
coset algebra (2.14). For the quantum coset $\slth_k / \slt$
we have checked that one  `would-be-generator' for each of
the scale dimensions 6, 8 and 9 becomes a null field at
$k=-{1 \over 2}$.
This probably also applies to the quantum analogue
of the classical generator at scale dimension 11.
Therefore, the first
classical relation, which arises at scale dimension 11, is
now not needed to
cancel a `would-be-generator', but can give rise to the first null
field (the one we already understood in the $\beta$-$\gamma$ realization).
By deformability, the same remarks apply to the quantum coset
(2.16). The picture in the classical case is less clear.
{}From the quantum equivalence in (1.21) and our study
of the classical analogue ${\cal P}_{sl(2)}$ of ${\hat {\cal P}}_{sl(2)}$,
we expect the classical limit
of the quantum coset  (2.16) to contain one generator for each positive even
scale dimension.
However, in the classical case
nothing happens to the generators at any value of the level $k \ne 0$,
and therefore  the set of  generators
of the `full classical coset' is
 encoded in the counting function (2.12) also at $k=-{1 \over 2}$.
This apparent
contradiction can probably be explained by the existence of a subring
closed under Poisson bracket, containing one generator for
each $d$, $0 < d \in 2 \Zed$, in the full classical coset
given by the ring of all classical invariants.
We expect the classical limit of (2.16) to realize such a proper subring
of the full classical coset.

Recall that at $k=1$ the quantum coset (2.14) gives just
the Virasoro algebra.
The classical limit of this Virasoro
field freely  generates a subring of the full classical
coset, which is obviously closed under the Poisson bracket.
(As noted above,  at $k=-{1 \over 2}$  we also expect
a subring, but an infinitely generated one, to realize the classical limit).
A similar remark applies to the cosets
$\GH_1 \oplus \GH_m / \GH_{m+1}$ for $\G$ any
simply laced Lie algebra $A$, $D$, $E$. The quantum versions
of these cosets are well-known to give rise to the so-called
`Casimir algebras' (see e.g.\ \q{\bbss}) which contain
no null fields for generic $m$.
On the other hand, at $m=\infty$
the ${\rm rank }\left(\G\right)$ classical Casimir invariants
(invariants without derivatives of the current)
freely generate a subring of the ring of all classical invariants,
which is closed  under Poisson bracket \q{\annals}.
Clearly, the deformation of this  subring to generic $m$ carries
the classical limit of the quantum Casimir algebra.

\bigskip
\centerline{\bf 3.\  On general cosets and
                     the deformable $\cW(2,3,4,5)$-algebra}
\medskip

In this section we give a prescription for finding the
generating set for a general class of (classical) cosets,
generalizing the results on the diagonal cosets (2.4)
discussed previously. In particular, this shows  that classical
cosets are infinitely, nonfreely generated in general.
These results will be illustrated
with another example which, as a by-product, also explains
the quantum $\cW(2,3,4,5)$-algebra found in \q{\hornfeck} in addition
to  the well understood Casimir algebra based on $A_4$
that has the same spectrum of generators.

Let $\cW$ be any classical $\cW$-algebra (including Kac-Moody algebras)
generated by finitely many, independent generating fields,
and suppose that $\cW$  contains a current algebra,  $\GH_\k$,
as a proper subalgebra. Suppose also
that the restriction of the central term of $\cW$
to $\GH_\k$ is nondegenerate, and that it is possible to
partition the generating fields of $\cW$
into the generating fields $J_a$ of $\GH_\k$
and a complementary set of generating fields  $J^\perp_i$ that
form  primary field multiplets with respect to the
current algebra $\GH_\k$. (To avoid confusion,
note that the $J^\perp_i$ need not be Kac-Moody currents).
Thus the Poisson brackets between the $J_a$  are
similar to those in (2.5), with a nondegenerate matrix $g_{ab}$,
except that now  we do not require the horizontal subalgebra
$\G\subset \GH_\k$ to be (semi)simple --- in particular,
it can contain $U(1)$ factors.
Moreover, we have Poisson brackets of the form
$$
\{ J_a(x), J^\perp_i(y)\} =-\sum_j R(a)_i^j J^\perp_j(y) \delta(x-y)\,,
\eqno(3.1)$$
where the matrices $R(a)=R(a)_i^j$, $a=1,\ldots, {\rm dim}\left(\G\right)$,
form a --- in general reducible ---  representation of $\G$,
$$
R(a) R(b) - R(b) R(a) = \sum_c f_{ab}^c R(c).
\eqno(3.2)$$
We are interested in the coset algebra
$$
{\cW\over \GH_\k},
\eqno(3.3)$$
{\it i.e.}, the Poisson bracket algebra carried
by the ring of those differential  polynomials
$$
p=p(J,\pa J, \ldots, \pa^m J, J^\perp, \pa J^\perp, \ldots, \pa^n J^\perp)
\eqno(3.4)$$
that Poisson commute with the charges
$$
Q(\epsilon) := \oint dx\, \epsilon^a(x) J_a(x)
\eqno(3.5)$$
for arbitrary test functions $\epsilon^a(x)$.

We shall show below that the elements of the coset (3.3)
are the {\it $\G$-invariant} differential polynomials of the type
$$
P=P(J^\perp, {\cal D} J^\perp,\ldots, {\cal D}^n J^\perp),
\eqno(3.6)$$
where ${\cal D}J^\perp$ is the covariant derivative defined by
$$
{\cal D}J^\perp_i := \pa J^\perp_i +{1 \over \k}
    \sum_{a,b,j} J_b g^{b a} R(a)_i^j J^\perp_j,
\eqno(3.7)$$
with $\sum_b g^{a b} g_{b c} = \delta_{ac}$.
In particular, the elements of the coset (3.3) can be written
as polynomials in  $J^\perp=\{J^\perp_i\}$ and  its covariant
derivatives. The variables in the argument of the r.h.s.\
of (3.6) belong to the  representation $R(a)$ of $\G$,
and $P$ must be invariant under the natural action of $\G$ on its arguments,
which means that  $P$ must Poisson commute with $Q(\epsilon)$
for {\it constant} $\epsilon^a$.
The significance  of this result is that
it reduces the problem of describing
the ring carrying the classical coset (3.3) to a standard
(although not necessarily easy)
problem in the invariant theory of the finite dimensional
Lie algebra $\G$.
Thus, applying the same general reasoning as for the diagonal
coset, we see that the differential ring of invariants (3.3)
must be infinitely, nonfreely generated in general.

For the proof we first rewrite the polynomial $p$ in (3.4)
as a (uniquely determined) polynomial
$$
P=P(J, DJ,\ldots, D^m J, J^\perp, \D J^\perp,\ldots, \D^n J^\perp)
\eqno(3.8)$$
in the new variables ${\cal D}^k J^\perp_i$, $D^k J_a$, where
the new derivative $D^k J$ of the $\G$-valued current $J$
is recursively defined by
$$
D^k J:= \pa D^{k-1} J +{1 \over \k} [J, D^{k-1} J]
\quad\hbox{with}\quad
D^0 J= J.
\eqno(3.9)$$
We are looking for $P$ such that
$$
\{ Q(\epsilon), P\} =0\,,
\eqno(3.10)$$
and the new variables are advantageous for computing the
Poisson bracket since they have simple Poisson brackets with
$Q(\epsilon)$:
$$\eqalign{
\{  \left({\cal D}^k J^\perp\right)_i , Q(\epsilon)\}&=
\sum_{j,a} \epsilon^a R(a)^j_i \left({\cal D}^k J^\perp\right)_j\cr
    \{ \left( D^k J\right)_a , Q(\epsilon) \}&=
[\epsilon , \left(D^k J\right)]_a - \k \,
    \left(D^k {\pa \epsilon}\right)_a\,\cr}
\eqno(3.11)$$
where $\epsilon = \{ \epsilon^a\}$ is the arbitrary $\G$-valued test
function and the derivative $D^k\left({\pa \epsilon}\right)$ is defined
similarly to (3.9),
$$
D^k \left(\pa \epsilon\right) := \pa D^{k-1} \left(\pa \epsilon\right) +
{1 \over \k} [J, D^{k-1} \left(\pa \epsilon\right) ]
\quad\hbox{with}\quad
D^0 \left(\pa \epsilon\right)= \pa \epsilon .
\eqno(3.12)$$
One easily verifies (3.11) by induction on $k$ (or recalls
it from Yang-Mills theory).
Because of the derivation property of the Poisson bracket,
$\{ Q(\epsilon), P\}$ is a   {\it linear} expression
in the algebraically independent parameters $\pa^k \epsilon^a$,
$$
\{ Q(\epsilon), P\} = \sum_{a,k} P_{a,k} \pa^k \epsilon^a\,,
\eqno(3.13)$$
and thus (3.10) requires the polynomials $P_{a,k}$,
which depend on the same variables as $P$, to vanish.
Let us now single out an index $a$ and look at $P_{a,k_{\rm max}}$,
where $k_{\rm max}$ is the highest value of $k$ that appears
in (3.13) for this particular $a$.
It is easy to see from the second line of (3.11) that the contributions
to $P_{a,k_{\rm max}}$ come from those monomials in
the polynomial $P$ that contain (various powers of) the
highest derivative of the type $\left(D^l J\right)_a$,
given by
$\left(D^{k_{\rm max}-1} J\right)_a$, which appears
in (3.8).
Since those monomials are algebraically independent,
$P_{a,k_{\rm max}}=0$ requires all of them to vanish.
This then immediately implies that $P$ in (3.8) cannot
depend on the variables $\left(D^l J\right)_a$ for any $a$
and $l=0,1,\ldots$, proving that $P$ must be of the form (3.6).
For such a $P$ the requirement that (3.10) must hold for
any test function
$\epsilon$ is equivalent to the same requirement for
constant $\epsilon$ since no derivative of $\epsilon$
appears in the first line of (3.11),
completing the proof of our characterization
of the elements of the coset (3.3).

It is easy to recover the results on the diagonal
coset on the l.h.s.\ of (2.4) from the above.
In that case $\GH_\k$, $\k=k+m$, is the diagonal
subalgebra in the denominator and $J^\perp$ becomes
$I$ in (2.6a).
It is also easy to specialize the general result
to classical cosets of the type
$\GH_k \oplus \HH_m / \HH_{i k + m}$, where $i$ denotes
the Dynkin index of the embedding ${\cal H} \subset \G$.
The generators  comprising
$J^\perp$ are in this case given by
the currents in  the complement
of $\HH_{i k}$ in $\GH_k$ together with the currents in $\HH_{i k}
\oplus \HH_m$ formed according to (2.6a).
{}From this one can derive the natural generalization of (2.4):
$$\lim_{m \to \infty}
{ {\widehat {\cal G}}_k \oplus \widehat {\cal H}_m \over
  {\widehat {\cal H}}_{ik+m}}
= {{\widehat {\cal G}}_k \over {\cal H}}.
\eqno(3.14)$$
It also interesting to note
that this type of coset
has subalgebras according to
$${\GH_k \over \HH_{i k}} \subset {\GH_k \oplus \HH_m \over \HH_{i k + m}}
\supset {\HH_{i k} \oplus \HH_m \over \HH_{i k + m}},
\eqno(3.15)$$
and our description applies to the full coset as well as
to these subcosets separately.
When dealing with  the coset ${\GH_k/\HH_{i k}}$
the variable $J^\perp$ becomes the component of the
$\G$-valued current generating $\GH_k$ that lies in the
orthogonal complement ${\cal H}^\perp$ in the decomposition
$\G={\cal H} \oplus {\cal H}^\perp$; thus explaining our notation.

The relations (3.14) and (3.15) are also valid in the quantum case.
The r.h.s.\ of (3.14) can be quantized by just normal ordering,
whereas for the quantum analogue of the l.h.s.\  this
holds at least in the limit $\hbar \to 0$.
It follows that for generic values of the parameters
there must be a one-to-one linear correspondence between
the fields in the classical and respective quantum cosets.
(It is the algebraic relations based on the ordinary product and
on the normal ordered product which  are different
in the classical and corresponding quantum case).
In particular, this means that for generic parameters
the number of invariant fields at a fixed scale dimension
in the  quantum coset (3.14) is equal to the number
of fields in the classical coset.
(Clearly, this applies to the general case (3.3) as well).
 This observation
can be utilized when calculating the vacuum character of a quantum coset
algebra for generic parameters.

Now we apply the above results to a particular class of classical
cosets with $\GH_\k = \widehat{U(1)}$. Consider a $\cW$-algebra
which can be divided into a
$\widehat{U(1)}$ subalgebra and a complement consisting of
$\widehat{U(1)}$-primary charge conjugate fields
with $U(1)$ charge $\pm 1$ and charge neutral fields. In this
case the polynomials (3.6) which are invariant under global
$U(1)$ transformations are the charge neutral combinations.
The charge neutral generators of $\cW$
simply survive the reduction and therefore we focus our attention
on the charged fields. Denote the charge conjugate
doublets by $W_a^\pm$ and their
covariant derivatives by $W_a^{\pm,(i)} = \D^i W_a^\pm$.
Then a redundant generating set for the coset is given by
the $\widehat{U(1)}$-primary charge neutral fields and the
following composite generators:
$$U_{a,b}^{i,j} := W_a^{+,(i)} W_b^{-,(j)}.
\eqno(3.16)$$
Forgetting about the action of the derivative $\D$,
all relations satisfied by the redundant set of generators (3.16)
are generated by the following quadratic relations:
$$U_{a,b}^{i,j} U_{c,d}^{k,l} - \epsilon_{b,c} \epsilon_{b,d}
\epsilon_{d,c} U_{a,d}^{i,l} U_{c,b}^{k,j} = 0,
\eqno(3.17)$$
where the `statistics factor' $\epsilon_{a,b}$ is defined
as $\epsilon_{a,b} = -1$ if both $W_a^{\pm}$ and $W_b^{\pm}$
are fermions and 1 otherwise.
 The proof that (3.17)  generates all relations
between the generators (3.16)
is a particularly simple version of a standard
argument used in invariant theory
(see the straightening algorithm in \q{\KR}).
 One has to show that a basis
for the linear space of invariant polynomials in the $W_a^{\pm,(i)}$ with a
given degree can be obtained from (3.16) using (3.17) and
(anti-)commutativity of the $U_{a,b}^{i,j}$. Because the invariant
monomials have zero charge they contain an equal number of $W_a^{+,(i)}$
and $W_a^{-,(i)}$, in particular they are even order. We can
choose an ordering where the charge alternates. This reduces
the problem to finding a basis for monomials of type
$W_a^{+,(i)} W_b^{-,(j)} W_c^{+,(k)} W_d^{-,(l)} \ldots$.
Clearly, a basis for these monomials is obtained by
a lexicographic ordering of the sets $\{(a,i),(c,k),\ldots\}$
and $\{(b,j),(d,l),\ldots\}$.
Then observe that, on the one hand, each such monomial
can be written as a monomial in the generators (3.16). On
the other hand, each set of indices can be independently
ordered for monomials in the $U_{a,b}^{i,j}$ using (3.17)
and (anti-)commutativity of the $U_{a,b}^{i,j}$. This completes
the proof.

Finally, we enforce the action of the covariant derivative
on the generators (3.16):
$$\D U_{a,b}^{i,j} = U_{a,b}^{i+1,j} + U_{a,b}^{i,j+1}.
\eqno(3.18)$$
Using (3.18) we can eliminate generators from the set (3.16)
in favour of a nonredundant set of generators,
for example the generators $U_{a,b}^{0,j}$.
As usual, the complete set of
relations satisfied by the nonredundant generating set $U_{a,b}^{0,j}$
can be obtained from (3.17) using (3.18).
(We note in passing that
the relation for the coset $W_3^{2} / \widehat{U(1)}$
discussed in \q{\ragoucy}, eq.\ (2.26) of the reference,
can be recovered
as a particular consequence of (3.17) and (3.18)).

The general statements will now be illustrated in the
following example: The commutant of the $U(1)$-current in the
$N=2$ super Virasoro algebra, SVIR($N=2$).
The algebra SVIR($N=2$) is
generated by the energy momentum tensor $L$, a current $J$
and two fermionic fields $G^\pm$ of scale dimension ${3 \over 2}$
carrying $U(1)$-charge $\pm 1$. In order to find the complement
of the $\widehat{U(1)}$ current algebra in SVIR($N=2$) we first go
to a primary basis with respect to the current algebra.
This means that we have to replace the energy momentum tensor $L$
by $\hat{L} = L - {3 \over 2 c} J J$ if we fix the normalization
such that the central term of $J$ with itself is given by
${c \over 3}$. Next we introduce the covariant derivative
according to (3.7):
$$\eqalign{
\D \hat{L} &:= \pa \hat{L} \ , \cr
\D G^{\pm} &:= \pa G^{\pm} \pm {3 \over c} \, J \, G^{\pm} \, . \cr
}\eqno(3.19)$$
{}From (3.16) and the action (3.18) of the covariant derivative $\D$
we obtain a nonredundant generating
set for the classical coset $\hbox{SVIR}(N=2) / \widehat{U(1)}$:
$$\eqalignno{
\hat{L} &= L - {3 \over 2 c} J J \ , &({\rm 3.20a}) \cr
\{ U^{0,j} &= G^{+} \D^j G^{-} \mid 0 \le j \in \Zed \}. &({\rm 3.20b}) \cr
}$$
Furthermore, we obtain
from (3.17) the complete
set of relations (in addition to (3.18))
satisfied by  the redundant set of generators
$U^{i,j} = (\D^i G^{+}) (\D^j G^{-})$:
$$U^{i,j} U^{k,l} + U^{i,l} U^{k,j} = 0.
\eqno(3.21)$$
An important special case of (3.21) for the nonredundant set
of generators (3.20b) is
$$U^{0,j} U^{0,k} = 0 \qquad \forall j,k.
\eqno(3.22)$$
These relations also directly follow from the fact
that $G^{\pm}$ are fermions satisfying the Pauli principle.
Note that the nonredundant set of generators (3.20) as well
as the set of relations they satisfy are infinite, as expected.

Let us now turn to the quantum version of this coset.
For generic $c$, counting the
number of invariant fields at a given scale dimension should
be the same at the classical and the quantum level.
Therefore one can
calculate to vacuum character, $\chi_0$, by counting the number of
classical invariants that arise as differential polynomial
in $\hat{L}$ and $G^{\pm}$.
In this way one obtains
$$\chi_0(q) - \phi_{2,3,4,5}(q) = -q^8 (
2 + 4 q + 9 q^2 + 16 q^3 + 32 q^4 + 54 q^5 +96 q^6 + {\cal O}(q^7)),
\eqno(3.23)$$
where $\phi_{2,3,4,5}$ denotes the vacuum character of a freely
generated algebra with generators of scale dimension 2, 3, 4 and 5.
This suggests the identification
$${\hbox{SVIR}(N=2) \over \widehat{U(1)}} \cong \cW(2,3,4,5)
\eqno(3.24)$$
for the quantum coset.
According to (3.23),
our quantum coset has two generic null fields at scale dimension 8.
Recall now that
the same null field structure
has been observed in \q{\hornfeck}
for the `second' deformable $\cW(2,3,4,5)$-algebra found by direct
construction.
Therefore the $\cW(2,3,4,5)$-algebra appearing in the identification
(3.24) must be the `second' $\cW(2,3,4,5)$-algebra of \q{\hornfeck}
(the other deformable $\cW(2,3,4,5)$-algebra
 is the Casimir algebra of $A_4$ which has no generic null field).

The identification  (3.24) is further supported
by explicit calculations presented in detail in \q{\ehhh}.
The energy momentum tensor of the quantum coset is just the
normal ordered version of (3.20a). The quantum analogues of the
generators of higher scale dimension (3.20b) are more difficult to
determine. By commutation with the zero mode of the current
it follows that all fields in the coset must
be uncharged. It is straightforward to make an ansatz in the
uncharged fields and determine those linear combinations that commute
with the complete current. One finds precisely one new generator
at scale dimensions 3, 4 and 5 with leading
terms quadratic in the fermions $G^{\pm}$,
and by computing the OPEs of these composite fields
one recovers \q{\ehhh} the
structure constants of the $\cW(2,3,4,5)$ given in \q{\hornfeck}.

Similarly as in previous examples, one can argue that normal ordered
analogues of the classical relations (3.22) ensure that the quantum
coset on the l.h.s.\ of (3.24) is
finitely generated, whereas the corresponding
classical coset needs infinitely
many generators (3.20). For example, the normal ordered version of the
classical relation $G^{+} G^{-} \  G^{+} G^{-} = 0$ contains
a term proportional to what would have been the new generator at
scale dimension 6.

The central charge $\hat{c}$ of the coset energy momentum tensor
$\hat{L}$  is obtained by
shifting the original central charge $c$  by 1.
Inserting the parametrization $c(k)$ of the unitary
minimal models of SVIR($N=2$) by a positive integer $k$, one
obtains
$$\hat{c}(k) = c(k) - 1 = {2 (k - 1) \over k+2}.
\eqno(3.25)$$
Note that this is just the well-known formula for the central charge
of $\Zed_k$-parafermions \q{\fatzam}. In fact, the coset algebra (3.24) is a
universal object \q{\ehhh} for the first unitary minimal models
of the Casimir algebras based on $A_{k-1}$ which describe
the $\Zed_k$-parafermions \q{\BBSS,\fatlyk}.
Moreover, it is also well known (see {\it e.g.}\ \q{\bouschou,\ravanini})
that the $\Zed_k$-parafermions can be realized using the coset
$\slth / \widehat{U(1)}$, and therefore also this coset algebra
should be a universal object for the $\Zed_k$-parafermions.
It can be verified that the standard character counting argument of
\q{\bouwknegt,\bouschou} indeed predicts a $\cW(2,3,4,5)$
(a finitely generated algebra) for the quantum coset
$\slth / \widehat{U(1)}$.

Because the above contradicts an earlier claim \q{\bakri}, we wish
to draw the reader's attention to the further evidence
in \q{\ehhh},  which shows  that the claim of \q{\bakri} that the quantum
coset $\slth_k / \widehat{U(1)}$ requires infinitely many generators
(one for each integer scale dimension
greater than or equal to two)
is incorrect as it stands. First,
it can be shown \q{\ehhh}    that $\slth_k / \widehat{U(1)}$
is isomorphic to $\hbox{SVIR}(N=2) / \widehat{U(1)}$, which we
have argued here to be finitely generated.
Second, it can be verified explicitly \q{\ehhh}
that the quantum coset $\slth_k / \widehat{U(1)}$
contains the $\cW(2,3,4,5)$ of \q{\hornfeck} as subalgebra.
Finally,
comparing the vacuum character\footnote{${}^{5}$}{
Note that the corresponding formula (4.14) in \q{\bakri} contains a
misprint: The exponent of $f(q)$ should be 2, not 3.
}
of this quantum coset, $\chi_0$ in (3.23),
to the number of composite fields in the
$\cW(2,3,4,5)$ determined in \q{\hornfeck} up to scale dimension $8$,
proves that no new generator appears in the coset at
scale dimensions 6, 7 or 8, contrary to the claim of \q{\bakri}.
Thus we expect that the coset $\slth_k / \widehat{U(1)}$  is
actually isomorphic to the  $\cW(2,3,4,5)$-algebra with
generic null fields constructed in \q{\hornfeck},
though no complete  proof is available at the moment.
We hope to present a complete proof
along the lines of the $\beta$-$\gamma$ example
and character considerations to arbitrary order
elsewhere \q{\RL}.

\bigskip
\centerline{\bf 4.\ Orbifolds of $\cW$-algebras}
\medskip

In this section we  show that orbifolds of $\cW$-algebras
behave very  similarly to the cosets discussed in the previous
two sections.
All known examples of orbifolds are generated finitely at
the quantum level, but
possess infinitely generated classical analogues with infinitely
many relations.
One  orbifold example that will  be discussed here  also turns up as
the $k=2$ special case of the coset (2.14).
Namely, the quantum version of the coset
$\left(\slth_2 \oplus \slth_m\right)/\slth_{m+2}$
leads to a $\cW(2,4,6)$-algebra  isomorphic to the bosonic
projection of the $N=1$ super Virasoro algebra \q{\bouwknegt}.
The structure of this quantum $\cW(2,4,6)$ has been investigated in
\q{\bouwknegt,\horst,\commute}. In particular, the
first generic null field appears at scale dimension
10. This bosonic projection is a particular
case of the more general `orbifold'-construction on $\cW$-algebras.

Any $\cW$-algebra (including Kac-Moody algebras) with nontrivial
outer automorphisms can be projected onto the invariant
subspace under the automorphism group. This so-called `orbifolding'
leads to another $\cW$-algebra. For simplicity,  we here restrict
our attention
to $\Zed_2$ automorphisms $\rho$ that act on the finitely many generators
$\{W_a \mid a \in \I \cup \K \}$ as follows:
$$\eqalign{
\rho(W_a) =& W_a \phantom{-} \qquad \forall a \in \K \ , \cr
\rho(W_b) =& -W_b \qquad \forall b \in \I \ . \cr
}\eqno(4.1)$$
We can divide the index set $\I$ into two subsets: A set
$\I_1$ referring to {\it bosonic} fields and a set $\I_2$
referring to {\it fermionic} fields transforming
nontrivially under the automorphism $\rho$.
It is easy to determine a generating
set {\it classically}. Note that the nontrivial $\rho$-invariant
differential polynomials are even order in the $\{W_b \mid b \in \I \}$.
Plainly, every even order polynomial can be regarded as a polynomial
in quadratic expressions. Therefore the quadratic expressions formed
out of the $\{W_b \mid b \in \I \}$ generate the orbifold together
with the invariant fields $\{W_a \mid a \in \K \}$. A redundant
set of quadratic generators is given by:
$$X_{b,c}^{i,j} := W_b^{(i)} W_c^{(j)}
\qquad b,c \in \I, \ 0 \le i,j \in \Zed
\eqno(4.2)$$
where $W_b^{(i)} := \pa^i W_b$. The derivative acts on the generators
(4.2) like in eq.\ (1.8):
$$\pa X_{b,c}^{i,j} = X_{b,c}^{i+1,j} + X_{b,c}^{i,j+1} .
\eqno(4.3)$$
Using the action of the derivative (4.3) and paying
attention to the Pauli principle for the fermionic generators,
 {\it i.e.}, that fermions have odd Grassmann parity, one can choose
the following minimal set of generators for the orbifold:
$$\eqalign{
W_a , & \ a \in \K \qquad \qquad \qquad \qquad \qquad \qquad \qquad \qquad
\ \ \hbox{(invariant generators)} \ , \cr
X_{b,c}^{0,j} &:= W_b \pa^j W_c , \quad b<c , \ b,c \in \I, \ 0 \le j\in\Zed
                 \ , \cr
X_{d,d}^{0,j} &:= W_d \pa^j W_d , \quad d \in \I_1, \ 0 \le j \in 2 \Zed
          \qquad \qquad \hbox{(square of bosons)} \ , \cr
X_{e,e}^{0,j} &:= W_e \pa^j W_e , \quad e \in \I_2, \ 0 < j \in 2 \Zed + 1
          \qquad \ \hbox{(square of fermions)} \cr
}\eqno(4.4)$$
where `$b<c$' denotes some ordering of the original generators. Eq.\ (4.4)
shows that $\Zed_2$ orbifolds are always {\it infinitely} generated at
the classical level. Note that a (redundant) set of quantum generators can be
obtained just by normal ordering (4.4).

In order to find the complete set of relations we first regard all
$W_a^{(i)}$ as independent. The complete set of relations satisfied
by the redundant set of generators (4.2) is generated by
$$\eqalignno{
X_{b,c}^{i,j} - \epsilon_{b,c} X_{c,b}^{j,i} &= 0, &({\rm 4.5a}) \cr
X_{b,c}^{i,j} X_{d,e}^{k,l} -
\epsilon_{c,d} X_{b,d}^{i,k} X_{c,e}^{j,l} &= 0, &({\rm 4.5b}) \cr
}$$
where
$\epsilon_{b,c} = -1$ if {\it both} $W_b$ and $W_c$ are
fermions, and $\epsilon_{b,c} = 1$ otherwise. (Clearly,
choosing certain indices in (4.5) equal leads to trivial relations).
The proof that (4.5) indeed generate all relations is very similar
to the one presented below (3.17) and therefore we omit it.

It is straightforward to derive the relations satisfied by the
nonredundant set of generators (4.4) from (4.5). One simply has to
recursively apply (4.3) (which encodes the action of the derivative) in
order to express the relations (4.5) in terms of the generators (4.4).

Next, we further elaborate some of these relations for two
examples and discuss their impact on the quantum case.
One of the simplest examples of orbifolds is the bosonic projection
of the $N=1$ super Virasoro algebra that we have already mentioned.
The $N=1$ super Virasoro algebra is the extension of the Virasoro
algebra $L$ by a primary scale dimension ${3 \over 2}$ fermion $G$.
According to (4.4),
a nonredundant set of generators for the classical orbifold is
$$L, \qquad \Phi^n := G \pa^n G \qquad \hbox{for all odd $n$}.
\eqno(4.6)$$
In particular, this orbifold has one generator at each positive
{\it even} scale dimension. Using the notation of (4.4) we have
the identification $\Phi^n = X^{0,n}$ if we omit the irrelevant
lower indices. From (4.5b) one reads off
$0 = X^{0,j} X^{0,l} + X^{0,0} X^{j,l} = X^{0,j} X^{0,l}$ because
$X^{0,0} = 0$. In terms of the generators (4.6) these
infinitely many relations read
$$0 = \Phi^n \Phi^m = 0 \qquad \hbox{for any} \quad 0 < n,m \in 2 \Zed + 1.
\eqno(4.7)$$
In this case the particular subset (4.7) of relations for
the nonredundant set of generators can also immediately be
inferred from the Pauli-principle:
$\Phi^n \Phi^m = (G \pa^n G) (G \pa^m G) = -G^2 (\pa^n G) (\pa^m G)
= 0$.
However, (4.5) encodes more relations. For example
(4.5) and (4.3) imply $X^{0,1} \pa^2 X^{0,1} =
X^{0,1} (X^{2,1} + 2 X^{1,2} + X^{0,3} )
= X^{0,1} X^{1,2} = -X^{0,1} X^{1,2}$. For the nonredundant
set of generators this implies the following relation at
scale dimension 10:
$$\Phi^1 \pa^2 \Phi^1 = 0.
\eqno(4.8)$$
It is also straightforward to verify (4.8) directly:
$\Phi^1 \pa^2 \Phi^1
= (G \pa G) \pa^2 (G \pa G) = G (\pa G) (\pa^2 G) (\pa G)
                          + 2 G (\pa G)^2 (\pa^2 G)
                          + G (\pa G) G (\pa^3 G)=0.$
The first and second term on the r.h.s.\ vanish because
$(\pa G)^2 = 0$, the last term is zero because $G^2 = 0$.

Now we explain how the classical relations make
the quantum version of the bosonic projection of the $N=1$
super Virasoro algebra generated by just three fields of scale
dimension 2, 4 and 6. First, one checks that the normal ordered
analogue of the classical relation $\Phi^1 \Phi^1 = 0$
contains a correction term proportional to $\Phi^5$
(for the explicit expression see (\appA.2) in appendix \appA). Thus,
the field $\Phi^5$ with scale dimension 8 does {\it not}
give rise to a new
generator in the quantum case and there is no relation
at scale dimension 8 in the quantum orbifold. Similarly,
the normal ordered counterpart of the classical relation
$\Phi^3 \Phi^1 = 0$ picks up correction terms containing
$\Phi^7$ (eq.\ (\appA.3) in appendix \appA).
This shows that also at scale dimension 10, upon normal
ordering, a classical relation cancels a `would-be-generator'.
It follows from general results on $\cW$-algebras that
this already
ensures that the quantum orbifold under discussion at least
contains a $\cW(2,4,6)$ as subalgebra. We have not yet used
the classical relation (4.8). Indeed, on the quantum
level there is a generic null field at scale dimension 10,
which is the normal ordered counterpart of this classical
relation (see eq.\ (\appA.4) in appendix {\appA} for an explicit formula).

So far one might have the impression that relations arise
in orbifolds just because of the Pauli principle. Note that
(4.5) also encodes infinitely many relations for orbifolds
of bosonic algebras. In order to illustrate
these relations and their impact also in the purely bosonic
case we briefly comment on a bosonic example.
Consider two commuting
copies of the Virasoro algebra ($L_1$ and $L_2$) with
equal central charge. Then $W := L_1 - L_2$ is primary with
respect to $L := L_1 + L_2$. Furthermore, $\rho(L) = L$
and $\rho(W) = -W$ is an automorphism of this $\cW(2,2)$.
According to (4.4) the subspace invariant under $\rho$
is generated, in the classical case,  by the following fields:
$$L, \qquad \tilde{\Phi}^n := W \pa^n W \qquad \hbox{for all even $n$}.
\eqno(4.9)$$
Again, we obtain one generator at each positive even scale
dimension. In this case rewriting the relations (4.5) for
the redundant set (4.3) in terms of the nonredundant set
(4.4) is slightly more complicated. Using the notation
$X^{i,j} = W^{(i)} W^{(j)}$ one checks that from (4.3)
and (4.5a)
$\pa^2 (X^{0,0} X^{0,0}) = 8 X^{0,1} X^{0,1} + 2 X^{0,0} \pa^2 X^{0,0}$
and $\pa^2 X^{0,0} = 2 X^{1,1} + 2 X^{0,2}$. Using (4.5b)
it is straightforward to check that
$\pa^2 (X^{0,0} X^{0,0}) + 6 X^{0,0} \pa^2 X^{0,0} + 8 X^{0,0} X^{0,2} = 0$.
This relation arises at scale dimension 10, which is the lowest scale
dimension admitting a relation. In terms of the generators (4.9) it reads
$$\pa^2 (\tilde{\Phi}^0 \tilde{\Phi}^0)
- 6 \tilde{\Phi}^0 \pa^2 \tilde{\Phi}^0
+ 8 \tilde{\Phi}^0 \tilde{\Phi}^2 = 0.
\eqno(4.10)$$
Turning now to the quantum case one can check that the normal ordered
counterpart of (4.10) is not identically zero, but contains correction
terms including the scale dimension 10 field $\tilde{\Phi}^6$ (the explicit
formula is eq.\ (\appA.6) in appendix \appA). This indicates that the
$\Zed_2$ orbifold of the quantum $\cW(2,2)$ is a $\cW(2,4,6,8)$ -- a
finitely generated algebra, which can be confirmed by an inspection
of its vacuum  character (see appendix \appA).

\bigskip
\centerline{\bf 5.\ A classical first class Hamiltonian reduction}
\medskip

In order to convince the reader that at the classical level
infinitely, nonfreely
generated algebras generically arise in {\it all} reduction procedures
applied to finitely, freely generated algebras, here we
present an example of a {\it first class} Hamiltonian reduction where
this is the case. Our starting algebra will be the DS type
${\cal W}_{\cal S}^{\cal G}$-algebra (see {\it e.g.}~\q{\FORT})
belonging to the $sl(2)$ embedding
${\cal S}$ associated to the long root of ${\cal G}=B_2$.
We next describe the structure of this
${\cal W}_{\cal S}^{\cal G}$-algebra.

The root diagram of the Lie algebra $\G=B_2$ consists of the vectors
$\pm e_1$, $\pm e_2$, $\pm (e_1\pm e_2)$,
and the algebra is spanned by  the Cartan-Weyl basis
$$
E_{\pm e_1},
\quad
E_{\pm e_2},
\quad
E_{\pm (e_1\pm e_2)}\,,
\quad
H_{e_1},
\quad
H_{e_2}\,,
\eqno(5.1)$$
normalized by $[H_{e_i},E_{e_i}]=E_{e_i}$.
We consider the $sl(2)$ subalgebra  $\S={\rm span\,}\{M_-,M_0,M_+\}$
belonging to the long root $(e_1+e_2)$,
$$
M_\pm:=E_{\pm (e_1+e_2)},
\quad
M_0 :={1\over 2}(H_{e_1}+H_{e_2})\,.
\eqno(5.2)$$
The  adjoint representation of $B_2$ decomposes under $\S$
as $10=3\times 1 +2\times 2 +3$ and the generating fields of
the $\cW_\S^\G$-algebra
are the components of the  `highest weight gauge' current,
$j_{\rm hw}(x)\in {\rm Ker}\left({\rm ad}_{M_+}\right)$, parametrized as
$$\eqalign{
j_{\rm hw}(x)=&I_+(z) E_{-(e_1-e_2)}+
I_0(x) (H_{e_1}-H_{e_2})
 +I_-(x)E_{e_1-e_2}\cr
&+{1\over 2} Z_+(x)E_{e_2} + {1\over 2} Z_-(z) E_{e_1}
 + {\cal L}(x)M_+\,.\cr}
\eqno(5.3)$$
The fields $I_{0,\pm }$ form an $\widehat{sl(2)}$ Kac-Moody subalgebra
of the $\cW_\S^\G$-algebra,
$$\eqalign{
\{ I_0(x), I_\pm(y)\}&=\pm I_\pm(y)\delta(x-y)\,,\cr
\{ I_0(x), I_0(y)\} &= {1\over 2} \kappa \delta'(x-y)\,,\cr
\{ I_+(x), I_-(y)\} &= 2I_0(x)\delta(x-y) +\kappa \delta'(x-y)\,,\cr}
\eqno(5.4)$$
where $\kappa$ is a nonzero constant.
The fields $Z_{\pm }$ are bosonic fields
with  conformal weight ${3\over 2}$
with respect to the Virasoro
$L:={1\over \kappa} \left( {\cal L} +I_- I_+ +I_0^2\right)$,
and form a doublet under the $\widehat{sl(2)}$ Kac-Moody subalgebra;
in particular they have $I_0$-charge $\pm {1\over 2}$.
This fixes almost all the Poisson brackets.
The Poisson brackets of $Z_\pm$ read
$$\eqalign{
\{ Z_\pm(x)\,,\,Z_\pm(y)\}&=\pm 2\kappa\left[ I'_\pm (y) \delta -
2I_\pm(y)\delta'\right] \,,\cr
\{ Z_-(x)\,,\,Z_+(y)\}&=2[{\cal L}-(I_-I_+ +I_0^2 +\k I_0')](y) \delta
+4\k I_0(y)\delta' -2{\k}^2\delta''\,.\cr}
\eqno(5.5)$$

We are interested in the classical
Hamiltonian reduction of this $\cW_\S^\G$-algebra defined
by the first class constraint
$$
I_+(x)=0\,.
\eqno(5.6)$$
The gauge group generated by this constraint
acts according to
$$\eqalign{
&L\longrightarrow L\,,\cr
&I_0\longrightarrow I_0\,,\cr
&Z_+\longrightarrow Z_+\,,\cr
&Z_-\longrightarrow Z_--\epsilon Z_+\,,\cr
&I_-\longrightarrow I_- -2\epsilon I_0 +\k \epsilon'\,,\cr}
\eqno(5.7)$$
where $\epsilon(x)$ is arbitrary.
The problem is just to describe the ring $\R$ of those differential
polynomials in the basic fields $L$, $I_0$, $Z_+$, $Z_-$, $I_-$
which are  invariant under the transformation rule (5.7).

By naive counting we expect that the reduced system should
have $4$ `functional degrees of freedom'.
We have the $3$ invariant fields $L$, $I_0$, $Z_+$ and can
easily construct a fourth invariant with the aid of the rational gauge
fixing implemented by putting $\epsilon={Z_-\over Z_+}$ in (5.7),
whereby
$$
I_-\longrightarrow R={B\over Z_+^2}
\quad\hbox{with}\quad
B:=Z_+^2 I_- - 2Z_+Z_-I_0 +\k (Z_-'Z_+ -Z_-Z_+')\,.
\eqno(5.8)$$
The scale dimension $4$ differential polynomial  $B$ is gauge invariant.
Why is our problem not completely trivial?
If we were looking for the differential {\it rational} gauge invariants
then $\{ I_0, Z_+, L,  R\}$ would clearly be a free generating
set.
However, since it is impossible to make sense
of quantum analogues of rational invariants,
we are interested in the differential {\it polynomial} invariants,
and, perhaps contrary to a naive expectation, the set
$\{ I_0, Z_+,L,  B\}$ is not a generating set for $\R$.
For example, the scale dimension $6$ invariant $K\in \R$
given by
$$
K:={{(\kappa B' Z_+' -\kappa BZ_+'' -2BI_0 Z_+')}\over Z_+}
\eqno(5.9)$$
cannot be expressed as a differential polynomial in the set
$\{ I_0, Z_+, L,B\}$, although it
can be checked to be a differential polynomial
in the basic fields.

Observe also that if we include $K$ into the generating set
of $\R$ then it will
contain the $5$ element subset $\{ I_0, Z_+, L, B, K\}$
subject to the  differential--algebraic relation
$$
Z_+ K -\kappa B' Z_+' +\kappa B Z_+'' +2 B I_0 Z_+'=0.
\eqno(5.10)$$
In a sense the problem is to find the higher
scale dimension analogues of the invariant $K$ (5.9) and
the relation (5.10).

It can be shown (for a proof, see appendix \appB) that
the ring $\R$ is generated by  the following gauge invariant
differential polynomials:
$$
L,\quad I_0,\quad Z_+,\quad P_{i,j},
\eqno(5.11)$$
where, for  arbitrary nonnegative integers $i$, $j$,
$$
P_{i,j}=\pa^i {Z}_+ \D^j {Z}_- -
\pa^j {Z}_+ \D^i {Z}_-
+{1\over \k} \sum_{a>0} \D^{a-1} {I_-}
\left( \pa^i {Z}_+ \pa^{j-a}{Z}_+  {j\choose a} -
\pa^{i-a} {Z}_+ \pa^j {Z}_+ {i\choose a} \right)
\eqno(5.12)$$
and $\D:=\pa -{2\over \kappa}I_0$ is a covariant derivative.
These generators of $\R$ are of course not independent.
They obey the fundamental nonlinear relations given by
$$
P_{i,j} P_{k,l} - P_{i,k}P_{j,l} +P_{i,l}P_{j,k}=0,
\eqno(5.13{\rm a})$$
$$
P_{i,j} \pa^k Z_+ - P_{i,k} \pa^j Z_+ + P_{j,k} \pa^i Z_+ =0,
\eqno(5.13{\rm b})$$
where $i,j,k,l$ are arbitrary nonnegative integers,
and the linear relations given by
$$\eqalign{
P_{i,j}  &=- P_{j,i} ,\cr
\D P_{i,j} & = P_{i,j+1} + P_{i+1,j}. \cr}
\eqno(5.14)$$
The reader may observe that (5.13-14) are
similar to the relations (1.6-8),
the reason for this is contained in
construction given in appendix \appB.

The generating set (5.11) is redundant on account of the linear
relations (5.14), and  using (5.14) like in section 1
we can find a minimal generating set, for instance
by keeping  only the generators $P_{1,2s}$ out of the $P_{i,j}$.
However, it is unavoidable that the minimal generating set of
$\R$ consists of {\it infinitely} many generators subject to {\it infinitely}
many relations. For completeness, we also note that the relation given
by (5.10) is recovered from (5.13b) by taking $i=0$, $j=1$, $k=2$ and
observing that $B$ in (5.8) is just $-\k P_{1,0}$, $K$ in (5.9)
is proportional to $P_{1,2}$, and  $P_{2,0}=\D P_{1,0}$.

We also wish to remark that the reductions related to the so-called
$W_{2n}^2$-algebras, which have been partially treated in the
appendix of \q{\FORT}, are very similar to the above example;
the invariant ring is infinitely generated in those cases, too.
The $W_{2n}^2$-cases and the above example are covered by the more general
model when one takes a Kac-Moody or $\cW$-algebra that contains
an $\widehat{sl(2)}$ Kac-Moody subalgebra and imposes a first class
constraint of the type (5.6).
We hope to further analyze  this class of classical reductions
and their quantum analogues in a future publication.

\bigskip
\centerline{\bf 6.\ Discussion}
\medskip

In this paper we pointed out  a new class of finitely
(but nonfreely) generated deformable quantum  $\cW$-algebras,
which consists of the algebras  possessing infinitely generated
classical limits obeying infinitely many differential--algebraic relations.
The existence of this class of $\cW$-algebras has not been noted previously,
although part of the results needed was already known in the literature.

Our main observation,  derived
from the examples, is that the infnitely many
relations satisfied by the classical generators
have the following twofold impact on the corresponding quantum
$\cW$-algebra. First, using a proper subset of the normal ordered
relations one can eliminate the infinitely many `would-be-generators'
in favour of a {\it finite} set of quantum generating fields.
Second, the generating fields in this finite set are still not
independent due to the rest of the normal ordered relations that
give rise to {\it generic null fields}.

These generic null fields
are nontrivial normal ordered differential polynomials in the
generating fields that exist for generic $c$ and
vanish identically on the defining vacuum representation
of the quantum $\cW$-algebra. As a terminological aside,
we propose to say that a finitely generated deformable
quantum $\cW$-algebra is {\it nonfreely generated} if such
generic null fields are present.

To be precise, we  fully demonstrated  the above
mentioned statements
only in the $\beta$-$\gamma$ example.
In the other examples we treated the first few relations
and confirmed by  character countings that the pairwise
cancellation of classical generators against classical relations
should happen upon normal ordering also at higher orders.
It is a very interesting open question whether this  mechanism
works to all orders in every case, or -- if not --
what are the conditions?
At least to our knowledge, no example (including $\slth / \widehat{U(1)}$
-- see section 3) is known for a quantum coset
(or other reduced algebra) obtained from a finitely generated
algebra for which  infinitely many generators are required
to describe the reduced quantum $\cW$-algebra.

Invariant theory (\q{\weyl-\KR}) played an
 important r\^ole in our considerations.
That this theory should be relevant for $\cW$-algebras
derives from the fact that, indeed, all known deformable
$\cW$-algebras can be obtained by reducing simple free field like
linear systems or affine Kac-Moody algebras,
and in reductions one is always interested in the invariants.
The class of invariants
most often considered in invariant theory is the polynomial
class and in our context the relevant class of invariants
is the {\it differential polynomial} class, which is not
very different. Since, as we have seen throughout the paper,
the generators and relations of the
differential ring of classical invariants arising in a reduction
know a lot about the generators and generic null fields of the
corresponding quantum $\cW$-algebra,
a more extensive application of invariant theory to $\cW$-algebras
should be a fruitful undertaking.

In this paper we emphasized that in a generic situation the
classical  coset, orbifold and first class reduction procedures
yield reduced classical systems carried by infinitely,
nonfreely generated differential rings of invariants.
(Incidentally,  as far as we know, the
classical coset and orbifold reductions lead to infinitely, nonfreely
generated classical systems without any nontrivial exception).
To avoid confusion, we should also stress that this
does not exclude the existence of a (possibly finitely generated)
subring closed under Poisson bracket
in such a way that only the subring is recovered from
the classical limit of a quantum $\cW$-algebra resulting from
the corresponding quantum reduction procedure.
In particular, if the quantum reduction
yields a $\cW$-algebra which is
generated finitely {\it without} generic null fields,
this is expected to be the case.
For instance, this happens for the coset realization
of `Casimir algebras' ($\cW$-algebras in the DS class for principal
embeddings), which
possess {\it finitely, freely} generated classical limits
(automatically obtained from classical DS reduction).
This provides us with an example where in the classical limit
of a quantum $\cW$-algebra resulting from a quantum (coset) reduction
one recovers only a finitely generated
subring of the full ring of  invariants resulting from
the analogous classical (coset) reduction.

We  wish to mention that in  \q{\FRS}
DS type classical $\cW$-algebras based on $B_n$ and $C_n$ have been
recovered from invariant subspaces of those based on some
$A_k$ by `folding'.
In our sense of `orbifolding' these subspaces are generated by
the invariant generators according to eq.\ (4.4), and thus
represent only a proper subring of the classical orbifold.
It should also be noted that
the quantum orbifolds of the Casimir algebras based
on $A_k$ in general have no subalgebras corresponding
to Casimir algebras based on a different simple Lie algebra
\q{\ehhh}.

Finally, we wish to point out that although the finitely but nonfreely
generated class of deformable quantum $\cW$-algebras considered in this
paper seems to be more complicated
than the already reasonably  well understood class of $\cW$-algebras
obtained from DS type reductions, it is also relevant for
`physical' applications.  For instance, a universal object for the
$\Zed_k$ parafermions (which have been widely used in the literature
--- see {\it e.g.}\ \q{\bouschou,\fatzam,\ravanini})
belongs to this class.  Furthermore,  in some cases
one might want not to include
any fermions into the symmetry algebra (for example because they are not
observable, or because -- from a mathematical point of view --
purely bosonic symmetry algebras are sometimes easier to handle),
which automatically leads to orbifolds belonging to this
new class.
The orbifolds obtained from bosonic
projection are the chiral algebras of GSO \q{\gso} projected
models that occur  in superstring theory.

\bigskip
\centerline{\bf Acknowledgements}
\medskip

We wish to thank R.\ Blumenhagen, W.\ Eholzer,
K.\ Hornfeck, R.\ H\"ubel,  M.\ R\"osgen, W.\ Nahm,
L. O'Raifeartaigh and I.\ Tsutsui  for
discussions and for reading the manuscript.
We are also grateful to P.\ Bouwknegt, A.\ Feingold and M.\ Scheunert
for correspondence and for pointing out useful references to us.
JdB was sponsored in part by NSF grant 9309888, and LF by the AvH Stiftung.

\vfill
\eject

\bigskip
\centerline{\bf Appendix \appA: Details on two quantum orbifolds}
\medskip

This appendix contains the normal ordered versions
of some classical relations for the orbifolds discussed in sect.\ 4.
The formulae below have been obtained on computer by using the
OPE package of \q{\thielemans}.

First we study the bosonic projection of the $N=1$ super Virasoro
algebra. Conventions are fixed by the following OPEs
of the energy momentum tensor $\Lh$ and the spin ${3 \over 2}$
fermion $\Gh$:
$$\eqalign{
\Lh(z) \Lh(w) &= \hb \left(
{c/2  \over (z-w)^4} + {2 \Lh(w) \over (z - w)^2  }
+ {\pa \Lh(w) \over z-w} \right) + {\rm reg.} \ , \cr
\Lh(z) \Gh(w) &= \hb \left(
{3/2 \Gh(w) \over (z - w)^2  }
+ {\pa \Gh(w) \over z-w} \right) + {\rm reg.} \ , \cr
\Gh(z) \Gh(w) &= \hb \left(
{2c/3 \over (z-w)^3} + {2 \Lh(w) \over z - w }\right) + {\rm reg.} \ . \cr
}\eqno({\rm \appA.1})$$
In order to be able to compare with the classical case we have
explicitly introduced the Planck constant $\hb$ in (\appA.1).
The normal ordered version of the simplest classical
relation $\Phi^1 \Phi^1 = 0$ becomes:
$$\eqalign{
((\Gh \pa \Gh) (\Gh \pa \Gh)) =
\hb \Biggl(&
{192 \hb - 31 c \over 90} (\Gh \pa^5 \Gh)
+ {7(c + 12 \hb ) \over 18} \pa^2 (\Gh \pa^3 \Gh)
- {7(2 c + 21 \hb ) \over 72} \pa^4 (\Gh \pa \Gh) \cr
&-{14 \over 3} ((\Gh \pa^3 \Gh) \Lh)
+ ((\Gh \pa \Gh) \pa^2 \Lh)
-3 (\pa (\Gh \pa \Gh) \pa \Lh)
+3 (\pa^2 (\Gh \pa \Gh) \Lh) \cr
&- \hb {1 \over 6} (\Lh \pa^4 \Lh)
- \hb {7 \over 3} (\pa \Lh \pa^3 \Lh)
+ \hb {41 c - 771 \hb \over 540} \pa^6 \Lh \Biggr). \cr
}\eqno({\rm \appA.2})$$
Apart from the quantized counterpart $(\Gh \pa^5 \Gh)$
of the classical generator $\Phi^5$ on the r.h.s.,  eq.\
(\appA.2) is a polynomial in the energy momentum tensor $\Lh$
and the dimension 4 and dimension 6 generators $(\Gh \pa \Gh)$
and $(\Gh \pa^3 \Gh)$. Therefore, $(\Gh \pa^5 \Gh)$ can
be eliminated and does {\it not} give rise to a new generator.
Note that the r.h.s.\ of (\appA.2) vanishes in the classical
limit $\hb \to 0$, as it should.
Similarly, normal ordering the classical relation $\Phi^3 \Phi^1 = 0$
leads to:
$$\eqalign{
((\Gh \pa^3 \Gh) (\Gh \pa \Gh)) =
\hb & \Biggl(
{128 \hb - 25 c \over 84} (\Gh \pa^7 \Gh)
+ {c + 11 \hb \over 2} \pa^2 (\Gh \pa^5 \Gh)
- {20 c + 161 \hb \over 24} \pa^4 (\Gh \pa^3 \Gh) \cr
&+ {60 c + 469 \hb \over 120} \pa^6 (\Gh \pa \Gh)
- {39 \over 10} ((\Gh \pa^5 \Gh) \Lh)
-6 (\pa(\Gh \pa^3 \Gh) \pa \Lh) \cr
&+5 (\pa^2 (\Gh \pa^3 \Gh) \Lh)
+ {1 \over 2} ((\Gh \pa \Gh) \pa^4 \Lh)
+ {7 \over 2} (\pa (\Gh \pa \Gh) \pa^3 \Lh) \cr
&- {5 \over 2} (\pa^4 (\Gh \pa \Gh) \Lh)
+ \hb {4 \over 5} (\Lh \pa^6 \Lh)
+ \hb {1 \over 5} (\pa \Lh \pa^5 \Lh)
- \hb {250 c + 6897 \hb \over 3360} \pa^8 \Lh \Biggr). \cr
}\eqno({\rm \appA.3})$$

\phantom{DIRTY TRICK}
\vskip 0.9 truecm

\noindent
Using (\appA.2) one can eliminate $(\Gh \pa^5 \Gh)$ in (\appA.3). Apart
from the term $(\Gh \pa^7 \Gh)$ both sides of (\appA.3)
are polynomials in the generators of scale dimension 2, 4 and 6.
This permits one to eliminate also $(\Gh \pa^7 \Gh)$, which
corresponds to the classical scale dimension 10 generator
$\Phi^7$.  Naturally, the r.h.s.\ of (\appA.3) vanishes in the
limit $\hb \to 0$.
Finally, the normal ordered counterpart of (4.8)
reads
$$\eqalign{
((\Gh \pa \Gh) \pa^2 (\Gh \pa \Gh)) =
\hb \Biggl( &
{8 (16 \hb  - c) \over 315} (\Gh \pa^7 \Gh)
+{c + 32\hb \over 20} \pa^2 (\Gh \pa^5 \Gh)
-{5 c + 6 \hb \over 12} \pa^4 (\Gh \pa^3 \Gh) \cr
&+{26 c + 93 \hb \over 90} \pa^6 (\Gh \pa \Gh)
- {16 \over 15} ((\Gh \pa^5 \Gh) \Lh)
- ((\Gh \pa^3 \Gh) \pa^2 \Lh) \cr
&-{8 \over 3} (\pa(\Gh \pa^3 \Gh) \pa \Lh)
+ {5 \over 3} (\pa^2 (\Gh \pa^3 \Gh) \Lh)
+ {1 \over 6} ((\Gh \pa \Gh) \pa^4 \Lh) \cr
&- {17 \over 6} (\pa (\Gh \pa \Gh) \pa^3 \Lh)
-3 (\pa^2 (\Gh \pa \Gh) \pa^2 \Lh)
- {5 \over 6} (\pa^3 (\Gh \pa \Gh) \pa \Lh) \cr
&- {3 \over 2} (\pa^4 (\Gh \pa \Gh) \Lh)
+ \hb {28 \over 45} (\Lh \pa^6 \Lh)
+ \hb {9 \over 5} (\pa \Lh \pa^5 \Lh) \cr
&+ \hb {7 \over 2} (\pa^2 \Lh \pa^4 \Lh)
+ \hb {89 \over 36} (\pa^3 \Lh \pa^3 \Lh)
- \hb {322 c + 13353 \hb \over 10080} \pa^8 \Lh \Biggr). \cr
}\eqno({\rm \appA.4})$$
After eliminating $(\Gh \pa^5 \Gh)$ from (\appA.2) and $(\Gh \pa^7 \Gh)$
from (\appA.3) the identity (\appA.4) is precisely the generic null field
at scale dimension 10 in the bosonic projection of the quantum
$N=1$ super Virasoro algebra.

\vskip 1.1truecm

The quantum version of the $\cW(2,2)$ obtained from two commuting
copies of the Virasoro algebra has the following OPEs:
$$\eqalign{
\Lh(z) \Lh(w) &= \hb \left(
{c/2  \over (z-w)^4} + {2 \Lh(w) \over (z - w)^2  }
+ {\pa \Lh(w) \over z-w} \right) + {\rm reg.} \ ,\cr
\Lh(z) \Wh(w) &= \hb \left(
{2 \Wh(w) \over (z - w)^2  }
+ {\pa \Wh(w) \over z-w} \right) + {\rm reg.} \ , \cr
\Wh(z) \Wh(w) &= \hb \left(
{c/2  \over (z-w)^4} + {2 \Lh(w) \over (z - w)^2  }
+ {\pa \Lh(w) \over z-w} \right) + {\rm reg.} \cr
}\eqno({\rm \appA.5})$$
where we have again  kept the Planck constant.
Now one can compute that the normal ordered counterpart of
eq.\ (4.10) is:
$$\eqalign{
\pa^2 ((\Wh \Wh&)  (\Wh \Wh)) - 6 ((\Wh \Wh) \pa^2 (\Wh \Wh))
+ 8 ((\Wh \Wh) (\Wh \pa^2 \Wh)) = \cr
\hb \Biggl(&
{\textstyle {47 c - 512 \hb \over 90}} (\Wh \pa^6 \Wh)
- {\textstyle {5 ( c + 32 \hb ) \over 4}} \pa^2 (\Wh \pa^4 \Wh)
+ {\textstyle {7 c - 74 \hb \over 6}} \pa^4 (\Wh \pa^2 \Wh)
+ {\textstyle {37 \hb - 7 c  \over 30}} \pa^6 (\Wh \Wh) \cr
&+ 20 ((\Wh \pa^4 \Wh) \Lh)
+ 36 ((\Wh \pa^2 \Wh) \pa^2 \Lh)
+ 28 (\pa(\Wh \pa^2 \Wh) \pa \Lh)
- 28 (\pa^2 (\Wh \pa^2 \Wh) \Lh) \cr
&- 4  (\pa (\Wh \Wh) \pa^3 \Lh)
- 20 (\pa^2 (\Wh \Wh) \pa^2 \Lh)
- {44 \over 3} (\pa^3 (\Wh \Wh) \pa \Lh)
+ 4 (\pa^4 (\Wh \Wh) \Lh) \cr
&+ \hb {6 \over 5} (\Lh \pa^6 \Lh)
- \hb {9 \over 5} (\pa \Lh \pa^5 \Lh)
- \hb {8 \over 3} (\pa^2 \Lh \pa^4 \Lh)
+ \hb {1 \over 6} (\pa^3 \Lh \pa^3 \Lh)
- \hb {\textstyle {184 c + 38895 \hb \over 5040}} \pa^8 \Lh \Biggr). \cr
}\eqno({\rm \appA.6})$$
Using (\appA.6) one can express $(\Wh \pa^6 \Wh)$ as a
polynomial in the generators $\Lh$, $(\Wh \Wh)$,
$(\Wh \pa^2 \Wh)$ and $(\Wh \pa^4 \Wh)$. This ensures
that the $\Zed_2$ orbifold of $\cW(2,2)$ has {\it no}
dimension 10 generator. Again, the r.h.s.\ of (\appA.6) vanishes
in the limit $\hb \to 0$ and it is not possible to eliminate
this generator at the classical level.

In the case of $\cW(2,2)$ we also would like to present a character
argument indicating that one does not need new generators at
higher dimensions either. Similarly to eq.\ (2.2), let $\phi_{2,4,6,8}(q)$
be the vacuum character of a $\cW(2,4,6,8)$ {\it without} relations.
It is straightforward to calculate the
character $\chi_0(q)$ of the submodule of the $\cW(2,2)$ vacuum
module invariant under $\rho$. Up to order $29$,  we
obtain for the difference of these two characters
$$\eqalign{
\chi_0(q) - \phi_{2,4,6,8}(q)
=- q^{12} \bigl(1 &+ 2 q + 5 q^2 + 9 q^3 + 29 q^5 + 53 q^6
    + 83 q^7 + 139 q^8 + 214 q^9 \cr
&   + 340 q^{10}
    + 510 q^{11} + 784 q^{12} + 1153 q^{13} + 1720 q^{14}
    + 2491 q^{15} \cr
&   + 3634 q^{16} + 5183 q^{17} + {\cal O}(q^{18}) \bigr). \cr
}\eqno({\rm \appA.7})$$
All coefficients are nonpositive
as expected. In particular, we read off from (\appA.7) that the
first generic null field appears at scale dimension 12.

For more details on quantum orbifolds we refer the interested
reader to \q{\ehhh}.
\vfill
\eject

\def\pa{\partial}

\def\G{{\cal G}}
\def\K{{\cal K}}
\def\R{{\cal R}}
\def\W{{\cal W}}

\def\S{{\cal S}}

\def\eps{\epsilon}
\def\k{\kappa}
\def\D{{\cal D}}

\def\e{\epsilon}

\medskip
\centerline{\bf Appendix \appB:
             The invariants in the Hamiltonian reduction example}
\medskip

In this appendix we show that the
generating set of the ring $\R$ of differential polynomials
in $L$, $I_0$, $Z_+$, $Z_-$, $I_-$ invariant under the gauge group (5.7)
is given by (5.11-14) as  stated in section 5.
We shall do this by a construction which reduces the problem of
finding the generating set of $\R$ to a problem in the invariant theory of
the group $SL(2)$.

Note first that  since $L$ does not mix with the other variables under
the transformation (5.7),  it can be factored out from the problem,
{\it i.e.},
$\R$ is generated by $L$ and the invariants depending on
$I_0$, $Z_+$, $Z_-$, $I_-$.
Similarly,  $I_0$ can also  be factored out
if we introduce the new variables
$$
\tilde{L}:=L,\quad \tilde{I}_0 :=  I_0, \quad \tilde{Z}_+ : = Z_+,
\quad \tilde{Z}_-  :=  GZ_-, \quad \tilde{I}_- :=  GI_-,
\eqno({\rm \appB.1})$$
where $G$ is given by
$$
G(x) :=\exp \left( -{2\over \k} \int^x dt \, I_0(t) \right).
\eqno({\rm \appB.2})$$
Indeed,  in terms of these  variables the
gauge transformation rule (5.7) becomes
$$
{\tilde L} \rightarrow  {\tilde L}, \quad
{\tilde I}_0  \rightarrow {\tilde  I}_0, \quad
{\tilde Z}_+  \rightarrow {\tilde  Z}_+, \quad
{\tilde Z}_-  \rightarrow {\tilde  Z}_- -{\tilde\eps} {\tilde Z}_+, \quad
{\tilde I}_-  \rightarrow {\tilde I}_- + \k \pa {\tilde\eps,}
\eqno({\rm \appB.3})$$
with
$$
\tilde{\eps}  :=  G\eps
\eqno({\rm \appB.4})$$
being arbitrary since $\e$ in (5.7) was arbitrary.

Let $P(L,I_0,Z_+,Z_-,I_-)$ be
an arbitrary differential polynomial in  the  original
variables and
$\tilde{P}(\tilde{L},\tilde{I}_0,\tilde{Z}_+,\tilde{Z}_-,\tilde{I}_-)$
an arbitrary differential polynomial in the new variables.
Decompose $P$ as a sum $\sum_kP_k$, where $P_k$ contains the terms of
degree $k$,  where we assign degree $1$ to $Z_-$ and $I_-$, and degree
$0$ to the other variables.
Decompose $\tilde P$ as $\tilde{P}=\sum_k\tilde{P}_k$ in the same way.
The map $F$ from tilded polynomials to untilded ones defined by
$$
F: \sum_k \tilde{P}_k(\tilde{L},\tilde{I}_0,\tilde{Z}_+,
\tilde{Z}_-,\tilde{I}_-) \mapsto
\sum_k G^{-k} \tilde{P}_k (L,I_0, Z_+, GZ_-,GI_-)
\eqno({\rm \appB.5a})$$
is invertibe, and  the inverse $F^{-1}$ is given by
$$
F^{-1}: \sum_k P_k(L,I_0,Z_+,Z_-,I_-) \mapsto
\sum_k G^k P_k(\tilde{L},\tilde{I}_0, \tilde{Z}_+, G^{-1}\tilde{Z}_-,
G^{-1}\tilde{I}_-).
\eqno({\rm \appB.5b})$$
This map naturally induces
a one-to-one map between the respective invariant differential
polynomials.

According to the above,  it is  enough to describe
the differential polynomial invariants
in the tilded variables ${\tilde I}_-$, ${\tilde Z}_+$, ${\tilde Z}_-$
under the transformation rule (\appB.3), which we now take
in its (equivalent) infinitesimal form,
$$
\delta {\tilde I}_-= \kappa \pa {\tilde \e},
\quad
\delta {\tilde Z}_+=0,
\quad
\delta {\tilde Z}_-=-{\tilde \e} {\tilde Z}_+.
\eqno({\rm \appB.6})$$
It is also convenient to introduce the notation
$$
I^{(l)}=\pa^l {\tilde I}_-,
\quad
\zeta_-^{(l)}:=\pa^l {\tilde Z}_-,
\quad
\zeta_+^{(l)}:=\pa^l {\tilde Z}_+,
\quad
\theta_l:=\pa^l {\tilde \e},
\eqno({\rm \appB.7})$$
and let $I$, $\zeta_+$, $\zeta_-$, $\theta$ denote the corresponding
infinite component vectors.
Purely algebraically,
the problem is to find the most general
polynomial ${\tilde P}(\zeta_-,\zeta_+,I)$ which is  is invariant under
$$
\delta I^{(l)} =\kappa \theta_{l+1},
\quad
\delta \zeta_+^{(l)}=0,
\quad
\delta \zeta_-^{(l)}=-\sum_{m\geq 0} {l\choose m} \zeta_+^{(l-m)} \theta_m
\eqno({\rm \appB.8})$$
for arbitrary $\theta$.
{} From computing the variation of ${\tilde P}$ using the chain rule,
${\tilde P}$ must satisfy
$$
\kappa {\pa {\tilde P}\over \pa I^{(m-1)} }-
\sum_{l\geq m} {l\choose m} \zeta_+^{(l-m)}
 {\pa {\tilde P}\over \pa \zeta_-^{(l)}}=0\,,
\qquad \left( m\geq 0,\quad  {\pa {\tilde P}\over \pa I^{(-1)}}=0\right).
\eqno({\rm \appB.9})$$
Consider now the decomposition of $\tilde P$
according to the different powers of $I$, given
by\footnote{${}^{6}$}{The new letter $Q$
is used since this decomposition  is different from that in (\appB.5).}
$$
{\tilde P}=\sum_{r\geq 0} Q_r
\quad \hbox{where}\quad
Q_r(\zeta_-,\zeta_+,\lambda I)=\lambda^r Q_r(\zeta_-,\zeta_+,I),
\eqno({\rm \appB.10})$$
which
leads to a refined form of (\appB.9), namely,
$$
\kappa {{\pa {Q}_{r+1}}\over {\pa I^{(m-1)}}} -
\sum_{l\geq m} {l\choose m} \zeta_+^{(l-m)}
 {{\pa {Q}_r}\over {\pa \zeta_-^{(l)}}}=0\,,
\quad (r=0,1,\ldots)\,.
\eqno({\rm \appB.11})$$
This implies that
every invariant polynomial $\tilde P$
is uniquely determined by its $I$-independent term $Q_0$.
On the other hand, taking $m=r=0$ in (\appB.11) we obtain that
$Q_0$ is subject to
$$
\sum_{l\geq 0} \zeta_+^{(l)} {{\pa Q_0} \over {\pa \zeta_-^{(l)}}}=0,
\eqno({\rm \appB.12})$$
and the point is that this  equation has a simple group theoretic meaning.

To see this take the standard action of the Lie algebra $sl(2)$,
with generators $E$, $H$, $F$, on the variables $\zeta_{\pm}^{(l)}$,
$$\eqalign{
\delta_E \zeta_-^{(l)}&=\zeta_+^{(l)},
\qquad
\delta_E \zeta_+^{(l)}=0,\cr
\delta_F \zeta_+^{(l)}&=\zeta_-^{(l)},
\qquad
\delta_F \zeta_-^{(l)}=0,\cr
\delta_H \zeta_+^{(l)} &=  \zeta_+^{(l)},
\qquad
\delta_H \xi_-^{(l)}=-\zeta_-^{(l)},\cr}
\eqno({\rm \appB.13})$$
and  extend this by the Leibniz rule to an action of $sl(2)$ on the
ring of polynomials in the infinitely many doublets $\zeta_\pm^{(l)}$.
Clearly, (\appB.12) just defines the subring of
`highest weight polynomials', {\it i.e.}, the kernel of $\delta_E$.
Using the representation theory of
$sl(2)$ it is not hard to see that the kernel of $\delta_E$
is generated by the pairwise symplectic scalar products of the different
doublets and the  highest weight components of the doublets themselves,
given by
$$
w_{i,j}:= \zeta_+^{(i)}\zeta_-^{(j)}-\zeta_-^{(i)}\zeta_+^{(j)},
  \quad \forall i\neq j,
\quad\hbox{and}\quad \zeta_+^{(l)}, \quad \forall\,l.
\eqno({\rm \appB.14})$$
(For instance, one can observe that the polynomials
depending on a finite subset of the doublets and having definite
degrees of homogeneity in those variables are an invariant subspace
and   using this one can show inductively that (\appB.14) indeed
generates the kernel of $\delta_E$).
One can also easily verify the relations
$$
w_{i,j}+w_{j,i}=0,
\eqno({\rm \appB.15})$$
$$
w_{i,j}w_{k,l}-w_{i,k}w_{j,l}+w_{i,l}w_{j,k}=0,
\eqno({\rm \appB.16})$$
$$
w_{i,j}\zeta_+^{(k)} - w_{i,k}\zeta_+^{(j)} + w_{j,k} \zeta_+^{(i)} =0,
\eqno({\rm \appB.17})$$
which are analogous to (1.6-7) in section 1.
Moreover, if we write
$$
\zeta_+^{(k)}=w_{k,\infty}=
     \zeta_+^{(k)}\zeta_-^{(\infty)}-\zeta_-^{(k)}\zeta_+^{(\infty)}
\quad\hbox{with}\quad
\zeta^{(\infty)}_-:=1,\ \zeta^{(\infty)}_+:=0,
\eqno({\rm \appB.18})$$
and let the indices $r$, $s$, $p$, $q$ run over
the nonnegative integers {\it and} $\infty$,
then we can uniformly write (\appB.15-17) as
$$
w_{r,s}+w_{s,r}=0,
\eqno({\rm \appB.19})$$
$$
w_{p,q}w_{r,s}-w_{p,r}w_{q,s}+w_{p,s}w_{q,r}=0.
\eqno({\rm \appB.20})$$
Then we can apply the `straightening algorithm'  given in Chapter 3 of
\q{\KR} to show that (\appB.19-20)  imply all the relations satisfied
by the generating set (\appB.14) of the ring of `highest weight polynomials'.
More precisely, this would be true if the variables $\zeta_\pm^{(l)}$
were independent, but now we have the derivation
$\pa \zeta_\pm^{(l)}=\zeta_\pm^{(l+1)}$,
which implies the extra linear relation
$$
\pa w_{r,s}=w_{r+1,s}+w_{r,s+1}.
\eqno({\rm \appB.21})$$

At this point
the solution space of (\appB.12) is fully under control
and  to derive the generating set of the gauge invariant differential
polynomials all one has to do now is to follow  the above construction
backwards. First one solves the recursion relation (\appB.11) taking any
of the elements in (\appB.14) for $Q_0$.
Using also (\appB.7),
this then yields the generating set of the invariants under (\appB.6).
Then one returns to the original variables by means of (\appB.5).
At the end of the day,  one obtains $P_{i,j}$ given by (5.12)
from $w_{i,j}$ in (\appB.14) by this procedure.
Moreover, one can trace back the relations given by (5.13-14)
to corresponding relations in (\appB.19-21).
Since it is completely straightforward from here on,
we omit the details of this derivation of the statement of section 5.

\vfill\eject

\bigskip
\centerline{\bf References}
\medskip
\baselineskip=12pt
\bibitem{\bouschou} P.\ Bouwknegt, K. Schoutens,
             {\it $\cW$-Symmetry in Conformal Field Theory},
             Phys.\ Rep.\ {\bf 223} (1993) p.\ 183
\bibitem{\bowwatts} P.\ Bowcock, G.M.T.\ Watts,
             {\it On the Classification of Quantum $\cW$-Algebras},
             Nucl.\ Phys.\ {\bf B379} (1992) p.\ 63
\bibitem{\dBT} J.\ de Boer, T.\ Tjin, {\it The Relation between Quantum
             $\cW$ Algebras and Lie Algebras}, preprint THU-93/05 (1993),
             IFTA-02-93, hep-th/9302006, to appear in Commun.\ Math.\ Phys.\
\bibitem{\FORT} L.\ Feh\'er, L.\ O'Raifeartaigh, P.\ Ruelle, I.\ Tsutsui,
             {\it On the Completeness of the Set of Classical $\cW$-Algebras
             Obtained from DS Reductions}, preprint BONN-HE-93-14 (1993),
             DIAS-STP-93-02, hep-th/9304125,
             to appear in Commun.\ Math.\ Phys.\
\bibitem{\fehort} L.\ Feh\'{e}r, L.\ O'Raifeartaigh, I.\ Tsutsui,
             {\it The Vacuum Preserving Lie Algebra of a Classical
             $\cW$-Algebra}, Phys.\ Lett.\ {\bf B316} (1993) p.\ 275
\bibitem{\kauwatts} H.G.\ Kausch, G.M.T.\ Watts,
             {\it A Study of $\cW$-Algebras Using Jacobi Identities},
             Nucl.\ Phys.\ {\bf B354} (1991) p.\ 740
\bibitem{\ehh} W.\ Eholzer, A.\ Honecker, R.\ H\"ubel,
             {\it How Complete is the Classification of $\cW$-Symmetries?},
             Phys.\ Lett.\ {\bf B308} (1993) p.\ 42
\footnote{${}^{7}$}{
Note that a remark on the classification of `exceptional' $\cW$-algebras
in this reference is slightly misleading because it was not recognized
that $\cW(2,8)$ at $c=-{712 \over 7}$
and $c=-{3164 \over 23}$ are related to minimal models of the
Casimir algebras of $E_8$ and $E_7$ respectively.}
\bibitem{\hornfeck} K.\ Hornfeck,
             {\it $\cW$-Algebras with Set of Primary Fields of Dimensions
             $(3, 4, 5)$ and $(3, 4, 5, 6)$},
             Nucl.\ Phys.\ {\bf B407} (1993) p.\ 237
\bibitem{\ragoucy} F.\ Delduc, L.\ Frappat, P.\ Sorba,
             F.\ Toppan, E.\ Ragoucy,
             {\it Rational $\cW$ Algebras From Composite Operators},
             Phys.\ Lett.\ {\bf B318} (1993) p.\ 457
\bibitem{\bakri} I.\ Bakas, E.\ Kiritsis,
             {\it Beyond the Large $N$ Limit: Non-Linear $\cW_\infty$
             as Symmetry of the $SL(2,\Real)/U(1)$ Coset Model},
             Int.\ Jour.\ Mod.\ Phys.\ {\bf A7}, Suppl.\ 1A (1992)
             p.\ 55
\bibitem{\weyl} H.\ Weyl, {\it The Classical Groups, Their Invariants
             and Representations}, Princeton, New Jersey, Princeton
             University Press (1946)
\bibitem{\howe} R.\ Howe, {\it `The Classical Groups' and Invariants
             of Binary Forms}, Proc.\ Symposia Pure Mathematics {\bf 48}
             (1988), p.\ 133;
             {\it Remarks on Classical Invariant Theory}, Trans.\ AMS
             {\bf 313} (1989), p.\ 539
\bibitem{\KR} J.\ Kung, G.-C.\ Rota, {\it The Invariant Theory of Binary
             Forms}, Bulletin of the AMS {\bf 10} (1984), p.\ 27
\bibitem{\bbss} F.A.\ Bais, P.\ Bouwknegt, M.\ Surridge, K.\ Schoutens,
             {\it Extensions of the Virasoro Algebra Constructed
             from Kac-Moody Algebras Using Higher Order Casimir Invariants},
             Nucl.\ Phys.\ {\bf B304} (1988) p.\ 348
\bibitem{\gno} P.\ Goddard, W.\ Nahm, D.\ Olive, {\it Symmetric Spaces,
              Sugawara's Energy Momentum Tensor in Two Dimensions
              and Free Fermions},
              Phys.\ Lett.\  {\bf B160} (1985) p.\ 111
\bibitem{\RL}  J.\ de Boer, L.\ Feh\'er {\it et al},  in preparation
\bibitem{\bouwknegt} P.\ Bouwknegt,
             {\it Extended Conformal Algebras from Kac-Moody Algebras},
             Proceedings of the meeting `Infinite dimensional
             Lie algebras and Groups',
             CIRM, Luminy, Marseille (1988) p.\ 527
\bibitem{\gosch} P.\ Goddard, A.\ Schwimmer, {\it Unitary Construction
             of Extended Conformal Algebras},
             Phys.\ Lett.\ {\bf B206} (1988) p.\ 62
\bibitem{\bogo} P.\ Bowcock, P.\ Goddard, {\it Coset Constructions
             and Extended Conformal Algebras},
             Nucl.\ Phys.\ {\bf B305} (1988) p.\ 685
\bibitem{\ralph} R.\ Blumenhagen, {\it $\cW$-Algebren in Konformer
             Quantenfeldtheorie}, Diplom\-arbeit BONN-IR-91-06 (1991)
\bibitem{\annals} J.\ Balog, L.\ Feh\'er, L.\ O'Raifeartaigh,
             P.\ Forg\'acs, A.\ Wipf,
             {\it Toda Theory and $\cW$-Algebra from a Gauged WZNW
             Point of View},
             Ann.\ Phys.\ 203 (1990) p.\ 76
\bibitem{\ehhh} R.\ Blumenhagen, W.\ Eholzer, A.\ Honecker, K.\ Hornfeck,
             R.\ H\"ubel, {\it Cosets and Unifying $\cW$-Algebras},
             BONN preprint, in preparation
\bibitem{\fatzam} V.A.\ Fateev, A.B.\ Zamolodchikov,
             {\it Nonlocal (Parafermion) Currents in Two-Dimen\-sional
             Conformal Quantum Field Theory and Self-Dual
             Critical Points in $\Zed_N$-Sym\-metric Statistical
             Systems},
             Sov.\ Phys.\ JETP {\bf 62} (1985) p.\ 215
\bibitem{\BBSS} F.A.\ Bais, P.\ Bouwknegt, M.\ Surridge, K.\ Schoutens,
             {\it Coset Construction for Extended Virasoro Algebras},
             Nucl.\ Phys.\ {\bf B304} (1988) p.\ 371
\bibitem{\fatlyk} V.A.\ Fateev, S.L.\ Lukyanov,
             {\it The Models of Two-Dimensional Conformal Quantum
             Field Theory with $\Zed_n$ Symmetry},
             Int.\ Jour.\ of Mod.\ Phys.\ {\bf A3} (1988) p.\ 507
\bibitem{\ravanini} F.\ Ravanini, {\it Informal Introduction to Extended
             Algebras and Conformal Field Theories with $c \ge 1$},
             preprint NORDITA-89/21P (1989)
\bibitem{\horst} H.G.\ Kausch,
             {\it Chiral Algebras in Conformal Field Theory},
             Ph.D.\ thesis, Cambridge University, September 1991
\bibitem{\commute} A.\ Honecker, {\it A Note on the Algebraic Evaluation
             of Correlators in Local Chiral Conformal Field Theory},
             preprint BONN-HE-92-25 (1992), hep-th/9209029
\bibitem{\FRS} L.\ Frappat, E.\ Ragoucy, P.\ Sorba,
             {\it Folding the $\cW$ Algebras},
             Nucl.\ Phys.\ {\bf B404} (1993) p.\ 805
\bibitem{\gso} F.\ Gliozzi, J.\ Scherk, D.\ Olive, {\it Supersymmetry,
             Supergravity Theories  and the Dual Spinor Model},
             Nucl.\  Phys.\ {\bf B122}  (1977) p.\ 253
\bibitem{\thielemans} K.\ Thielemans, {\it
             A Mathematica Package for Computing
             Operator Product Expansions},
             Int.\ Jour.\ Mod.\ Phys.\ {\bf C2} (1991) p.\ 787

\bye